\documentclass[10pt]{article}
\usepackage[margin=1.0in]{geometry}
\usepackage{amsmath,amsfonts}
\usepackage{algorithmic}
\usepackage{array}
\usepackage{comment}
\usepackage{caption}
\usepackage{enumerate}
\usepackage[shortlabels]{enumitem}
\usepackage{float}
\usepackage{graphicx}
\usepackage[acronym]{glossaries}
\usepackage{multirow}
\usepackage{subfigure}
\usepackage{varwidth}
\usepackage{wrapfig}
\usepackage{url}
\usepackage{xcolor}

\newcommand{\DD}{\mathcal{D}}

\newcommand{\LL}{\mathcal{L}}

\newcommand{\BL}[1]{\begin{list}{}{\leftmargin #1 \labelsep 0.2in}}
\newcommand{\EL}{\end{list}}
\newcommand{\BA}{\begin{array}}
\newcommand{\EA}{\end{array}}
\newcommand{\EQ}{\vspace{-0.05in}\begin{equation}}
\newcommand{\EN}{\vspace{-0.05in}\end{equation}}
\newcommand{\EQu}{\vspace{-0.05in}\begin{displaymath}}
\newcommand{\ENu}{\vspace{-0.05in}\end{displaymath}}
\newcommand{\EQa}{\begin{eqnarray}}
\newcommand{\ENa}{\end{eqnarray}}
\newcommand{\EQua}{\begin{eqnarray*}}
\newcommand{\ENua}{\end{eqnarray*}}

\newcommand{\qs}{\hspace*{1.5em}}
\newcommand{\hqs}{\hspace*{0.75em}}

\newcommand{\nn}{\nonumber\\}
\newcommand{\hh}{\\ \hline}

\renewcommand{\epsilon}{\varepsilon}

\definecolor{darkgreen}{rgb}{0.1, 0.6, 0.1}
\definecolor{lightblue}{rgb}{0.5, 0.5, 1.0}

\newcommand{\aeq}{&\!\!\!\!=\!\!\!\!&}
\newcolumntype{P}[1]{>{\centering\arraybackslash}p{#1}}

\title{Performance Health Index for Complex Cyber Infrastructures}
\author{Sanjeev Sondur and Krishna Kant\\ Temple University}

\begin{document}
\maketitle
\begin{abstract}
Most IT systems depend on a set of configuration variables (CVs), expressed as a name/value pair that collectively define the resource allocation for the system. While the ill-effects of misconfiguration or improper resource allocation are well-known, there is no effective a priori metric to quantify the impact of the configuration on the desired system attributes such as performance, availability, etc. In this paper, we propose a \textit{Configuration Health Index} (CHI) framework specifically attuned to the performance attribute to capture the influence of CVs on the performance aspects of the system. We show how CHI, which is defined as a configuration scoring system, can take advantage of the domain knowledge and  the available (but rather limited) performance data to produce important insights into the configuration settings. We compare the CHI with both well-advertised segmented non-linear models and state-of-the-art data-driven models, and show that the CHI not only consistently provides better results but also avoids the dangers of pure data drive approach which may predict incorrect behavior or eliminate some essential configuration variables from consideration. 

\end{abstract}

\section{Introduction}

As the data centers grow in complexity, sophistication, and size of the infrastructure and services supported, their proper configuration is becoming a huge challenge. Most objects from services down to virtual and physical devices have many configuration parameters (or variables), whose correct setting is crucial for proper functioning and good performance. Many state of art literature ~\cite{xu2016early,Zhou-CM-survey,AC-misconfig} highlight that  70\%-85\%  of  all  users’  configuration errors account for the high cost of misconfiguration. Added to this is the poor understanding (and miscommunication) of configuration variables\footnote{Most commonly referred to as features~\cite{guo2013variability,kang1990feature}} (CVs) on the ``outcome'' or behavior of the service/system, and hence results in the wrong setting or misconfigured options. Poorly configured systems (or resource allocation)  may fail to satisfy the performance, availability, security, and other goals and result in avoidable operational costs and user dissatisfaction. Misconfigurations are routinely exploited by attackers to gain entry or disrupt the system~\cite{config-errors}.  Ill-effects related to system misconfiguration are well documented~\cite{Zhou-CM-survey, icin2020sondur}, including their impact on the economy, security incidents, service recovery time, loss of confidence, social impact, etc. 

In Cloud computing applications, \textit{configuring} the right resources to the Cloud computing objects (i.e. Cloud storage, virtual machines, etc.) is becoming critical, both because of the complexity involved in allocating the right resources and understanding their overall effect on the system (e.g. cost of resource provisioning, user experience,  performance, energy consumption, etc.)~\cite{moradi2021online,Masanet2020,Makrani2021}. 
\textit{Resource provisioning} for Cloud based applications involves several unique challenges, wherein a Cloud instance is characterized (and priced) based on resource ``configuration'' (i.e. CPU family/cores, memory, and disk capacity)~\cite{wei2015towards,Zaman2013,Masanet2020,Wang2019}.   
%The problem of configuration optimization, i.e. optimize the \textit{behavior} of a system while minimizing an objective (e.g. cost) has received considerable interest~\cite{alipourfard2017cherrypick,zhu2017bestconfig,yadwadkar2017selecting}.
%The ``behavior'' of most entities in a cyber infrastructure, from devices to the entire infrastructure (and similarly from device drivers to user-level services running on the infrastructure) depends on its \textit{configuration}. 
%Configurations are user-settable parameters (most commonly referred to as features~\cite{kang1990feature}) that define how a resource is allocated (e.g. CPU cores, memory, OS threads, page size, etc), transactions processed, security applied, etc. Finding the right configuration of a system is important for users  (system administrators, designers, developers) as it determines the operational efficiency of the system. To achieve this, users need to understand the influence of various configuration values, (henceforth denoted as CV) on the system ``behavior'', which is predominantly expressed as \textit{performance}~\cite{westermann2012automated} (e.g., response time or throughput).

Real systems may have 10s to 1000s user-settable configuration variables (or CVs)~\cite{xu2015hey,krishna2020conex}; in addition, there could be a significant number of hidden or latent manufacturer provided parameters that are not well described. Each of these may take anywhere from two values (for binary variables) to an uncountable number of values, although in practice the feasible values may be limited to either an explicit set of values (e.g., installed memory being one of 32GB, 64GB, 96GB, and 128GB), or approximated by a number of buckets of potentially varying width. Even so, the configuration space (henceforth denoted as $\Omega$) quickly becomes too large to comprehensively characterize it. For example, 10 CVs with 10 values each, amount to $10^{10}$ or 10B combinations. 

Thus, we need a more compact way of {\em understanding the contribution of  the individual CVs} on the overall system performance in the context of other settings. For example, when deciding how much memory to put on a given web server, it is helpful to know roughly at what point the diminishing returns\footnote{The point beyond which, any additional resource allocation is detrimental to performance.} kick in sufficiently strongly to make the additional memory of dubious value. One could ask a similar question regarding the page size for a database or the local storage allocation for a Cloud Storage Gateway (CSG).  Since the main difficulty in evaluating configurations is the interaction among settings of different CVs ~\cite{xu2015hey,krishna2020conex,velez2020configcrusher,Wang2019}, we need a way of capturing the interactions in a compact manner. Configuration settings representing physical resources (e.g., computing cores, page-size) also relate to the cost constraints as well if we open up the possibility of expanding the existing systems; however, we do not address this aspect. In other words, there are predefined ranges for all CVs and any selection must stay within those limits. 

The work reported here is motivated by our earlier work in CHeSS~\cite{icin2020sondur} (Configuration  Health  Scoring  System), that defined a {\em scoring system} to compute the Configuration Health Index (CHI) for compactly assessing the influence of CVs on various attributes of a system (including performance, availability, security, etc.). We discuss the general CHI concept briefly in section~\ref{s:chiframework}. 
The difference from our earlier work is that, in this paper, we focus entirely on the performance related CHI and propose a way of exploiting the limited observational data and the domain expertise in order to robustly quantify the CHI scores. We discuss the importance of such an approach for production systems in section~\ref{s:domain}.

To demonstrate the merits of our approach, we use real world  configuration (public domain) data sets from a number of very different systems ~\cite{nair2017using, siegmund2015performance,TUDelftBitBrains}, and our study of the Cloud Storage Gateway ~\cite{Cloud2019Sondur}.  We show that with the same amount of available data, our method can produce significantly better results (e.g. health score of CVs, better prediction accuracy, and low variance) since we use data to estimate some key parameters, rather than the actual behavior itself.

The main contributions of this paper are as follows:
\begin{itemize}
    \item Define an \textit{a priori} mechanism for evaluating the quality of configuration of a service in form of a scoring system for its performance. 
    \item Demonstrate how the domain expertise can be exploited to yield more robust score quantification without overburdening the experts.
    \item Demonstrate that such a scoring system performs better than the state of the art techniques for a variety of configuration data sets used. 
\end{itemize}

\section{Configuration Health Scoring System}
\label{s:PHI-CHeSS}

The "health" of the system can be characterized along several dimensions (or {\em attributes}) such as security, availability, manageability, performance, etc. For each attribute, we need a measure that is generally considered indicative of that attribute without necessarily having to define a very specific measure, since specificity, while desirable, narrows the applicability of the health measure.

\subsection{Quantification of Health Score}

The scoring system must provide a compact characterization of the influence of {\em configuration variables (or CVs)} on various attributes of interest. A preliminary scoring system called CHeSS is presented in~\cite{icin2020sondur}. A score is a number between some lower bound \& upper bound (e.g. 0..2) where a mid-score (e.g. 1.0) corresponds to a nominal (or "average") configuration, the upper bound corresponds to a highly optimized configuration, and  the lower bound corresponds to a very poor (but still operational) configuration.  The purpose of a scoring system is to rate configurations in terms of a normalized measure of each attribute in a simple way in order to assess the health of the system as a function of the configuration parameters. The health index is not synonymous with very specific or detailed measures that require a detailed quantitative model; instead, it is intended as a measure over the configuration space that provides some indication of how good the configuration is.  The distinction is subtle. On one hand, we do want the score to reflect a suitable measure of the attribute (e.g., performance measured in terms of throughput, latency, and other important aspects); on the other, reducing the score to be simply a scaled version of the throughput is of little value, since it too will require detailed modeling. We want to avoid the need for detailed modeling in the context of configuration health because not only it requires a very specific measure, but it also is generally intractable due to a large number of configuration parameters, their interdependencies, and their complex influence on the chosen measure. 

Because of how the scoring system is targeted, it necessarily carries some level of nonspecificity both in the measures and the values. In particular, with a mid-score (e.g. 1.0) considered as a nominal score, it is the significant deviation from the mid-score that is important, not the precise value. The simplicity in defining and evaluating the score is crucial for scalability in dealing with large configuration spaces. Such nonspecificity is inherent to any scoring system, in particular, the well-known Configuration Vulnerability Scoring System (CVSS)~\cite{cvss-main} that has been used by the security community for quite some time and was the origination motivation for CHeSS. Note that CVSS scores are assigned entirely manually based on the "domain knowledge" which consists of both observed and expected impact of a security vulnerability.

\subsection{Configuration Specification}

Configurations are generally specified as name-value pairs defined in configuration files and stored in service specific format (e.g. json, xml, text file or local/remote repository).
Service functionality is an abstract term that can take various forms based on users or context,  either for processing data, securing services, energy consumed, etc.  An a priori scoring of the configuration, as envisioned by CHI, will aid the user-community (administrators, designers, developers, end-users, etc.) to gain an insight into the strength or weakness of the configuration beforehand, and hence minimize any costly after-facts.

A configuration file for the service includes a set of configuration "objects" and their settings. Configuration objects are often organized as a hierarchy, with a top-level object representing a feature (QoS, VPN, or VLAN in a router) or component (e.g., namenode \& datanode in HDFS), with lower level objects breaking it into finer aspects. For example, a router would have top-level objects for configuration settings of Layer2, Layer3, possibly Layer4, Security, Authentication, etc., each of which has further objects down below. For example, the Layer2 setting includes the spanning-tree protocol setting, with various VLAN settings under that, etc. Configuration settings serve a variety of purposes. Many of them are used for purposes other than the "health"; for example, the EXT4 file system has options for folding the case in directory searches, enabling extended attributes, support for huge files, etc.  Others affect the service health directly in terms of some attributes, e.g., enabling encryption that affects security and performance. However, the health impact of many configuration settings is not obvious and requires varying levels of domain knowledge to assess, e.g., in EXT4, enabling metadata checksum increases resilience, and enabling extent trees results in better performance. 

As an example, in CHeSS~\cite{icin2020sondur}, we considered routers in a commercial data center with complex and elaborate working configurations. The largest configuration file here had 22,000 lines and operating on an object hierarchy up to 7 levels deep. Thus the exercise  provides a good insight into the usefulness of a scoring system for  complex systems where detailed observational data is often spotty or simply unavailable (e.g., the impact of key length on security) and detailed quantitative modeling  difficult.  Given the intricacies of routing protocols and complex features involving VLANs, authentication, etc., we believe that weights assigned by highly experienced administrators can be regarded as good a depiction of "ground truth" as one might reasonably obtain in such an environment.  

A configuration file $c$ of a service contains several CVs henceforth denoted as $P_m,  1\le m \le M$. Each CV is a tuple representing the name and value pair ($P: p$). Depending on the name/value tuple, the configuration object can contribute to one or more health attributes of the service. For example in Fig.~\ref{f:sampleData-csg}, a configuration file $c_4$  contains a configuration object $P_1: \{mem=32\}$, which states that 32GB of memory is allocated. This statement can contribute towards the performance attribute by factor $p_1$ and security attribute by factor $s_1$. Similarly, another configuration object $P_2: \{cores=4\}$ may state CPU resource as 4 cores and contribute towards performance attribute as $p_2$ and availability attribute as $a_2$. Thus, each configuration object $P$ influences the service behavior and contributes to one or more attributes (denoted as $\vec{h}=\{p,s,a\cdots\}$).  The  goal of this research to identify these unknowns (i.e. $p_1$, $p_2$, $s_1$, $a_2$, etc.) based on the observable behavior of the service with the configuration file (e.g. observed performance in Fig.~\ref{f:sampleData-csg}, $O_1$=112Kbps).

\subsection{Challenges in Assigning Scores}
\label{s:configfile}

The key problem in defining CHI is twofold: (a) estimation of scores (or CHI values) for leaf-level objects in the configuration object hierarchy, and (b) composition of the scores along the hierarchy to determine a score of any arbitrary object.  Here (a) can range from a direct assignment of a score by a knowledgeable user/administrator up to an entirely automated estimation. We discuss this aspect in some detail, starting with an entirely manual assignment. We also discuss the composition method used in CHeSS and continue to use the same here as well.

\subsection{How can CHI help? Some preliminary work}

In CHeSS, we focus on CHI in general and evaluated a concrete example relative to three attributes, namely availability, security, performance, for a large commercial routing network. Since the impact of configuration parameters on attributes like availability or security is difficult to determine experimentally, the assignment of scores (or weights) to the leaf-level objects was done by experienced router administrators and then aggregated to estimate the weight or score of an object. This was done recursively from leaves to the root, the end result being the overall score for the router. However, it is not just the overall score, but intermediate scores that are also important in assessing the quality of the settings.   The aggregation was done using simple geometric means. As is well known, geometric mean preserves relative scaling (since geometric mean of ratios is the ratio of geometric means) and is tolerant of occasional erroneous weight assignment in a large hierarchy. The geometric means provide a measure of the \textit{configuration health index (CHI)} at each level and ultimately for the top-level objects. 

\begin{figure}[htb]
%\vspace{-6pt}
\centering
\includegraphics[width=0.75\columnwidth]{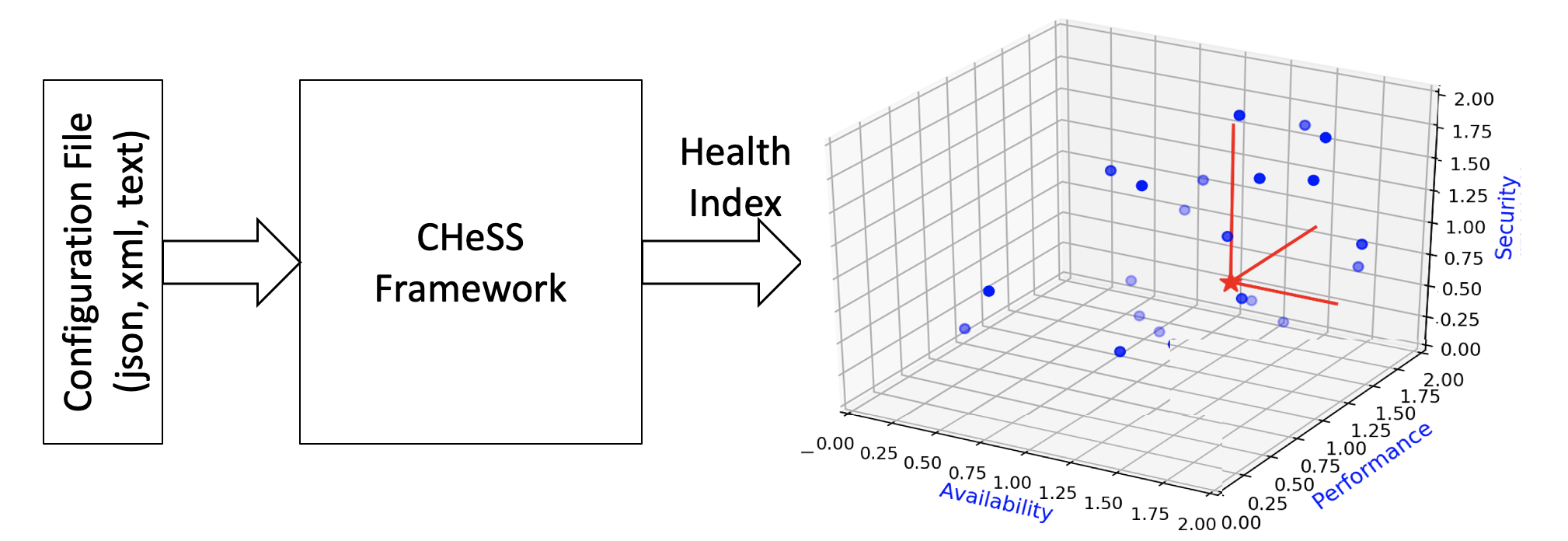}
\vspace{-3pt}
\caption{{CHeSS Framework{~\cite{icin2020sondur}}}}
\label{f:chess}
\vspace{-6pt}
\end{figure}

Given the weight, we express the Health Index (\textit{H}) metric of a configuration as a vector of impacted attributes. As shown in Fig~\ref{f:chess}, the framework takes the configuration file as input, analyses the configuration statements (CVs, aka configuration objects) for their influence on  different attributes, and quantifies the \textit{H} metric at all levels of the object hierarchy. The right hand side of Fig~\ref{f:chess} illustrates a sample result pictorially that we obtained for router configurations in a real data center. Here the vector \textit{H} consists of only three attributes $P,A,S$ (performance, availability, security). A point closer to the upper bounds indicates a good configuration, and different blue dots correspond to different highest level configuration objects of the same router. The figure clearly shows that some objects are quite poorly configured, especially with respect to availability and security. Such a depiction clearly indicates the value of CHI concept and allows the administrators to focus on objects whose configuration needs to be improved.

\subsection{Aggregation of CHI Scores}

For the rest of this paper, we will associate $\vec{H}$ with a single attribute representing performance $\mathbb{P}$ (performance is shown as an important attribute by Westermann et al.~\cite{westermann2012automated}), and health index ($\vec{h}$) of individual configuration object ($P$) is marked as a single metric $h$.
The overall metric (or quality of each attribute in \textit{H}) is represented as the geometric mean of all the contributing attributes (weights) $\vec{h_i}$'s from all the configuration objects $P_i$. The $H_n$ of the  configuration file $c_n$ (and hence the service) is then given as:

\EQ
H_n = \sqrt[{M}]{\left(\prod _{m=1}^{M} (h_{nm})\right)}  \nn
\label{e:HI}
\EN
or alternatively,

\EQa
log(H_n) \aeq \frac{1}{M} \sum_{m=1}^{M} log(h_{nm})
\label{e:logHI}
\ENa

\subsection{Exploiting Domain Knowledge for Performance CHI}
\label{s:domain}

The key challenge in such an approach \textit{is} to estimate the values of $h_i$’s. With most attributes, including security, availability, manageability, etc., it is generally infeasible to set the CVs to desired values and experimentally determine their impact. Instead, one must estimate the impact via some mathematical model calibrated based on some basic data that might be available. For example, availability (or reliability) modeling generally uses a simple compositional model based on the availability of individual components. Similarly, we may have some quantification of the attack probabilities, which along with suitable attack graph models can give us a quantification of the security of the system. The performance attribute is somewhat unique in this respect in that it is possible (at least in theory) to set the CVs to some values and measure the performance (or compute it based on a model calibrated from the observed behavior). This brings in the possibility of at least a partially data-driven determination of the scores, and thereby reduces the amount of effort required on part of the domain experts. 

However, we cannot immediately swing to the other extreme and claim that there is no need for domain expertise, and everything can be done in a purely data-driven manner. In fact, there are numerous hurdles in making a data-driven approach work, and we show in this paper that it can often lead to misleading results; instead, an approach that judiciously uses expert input can not only improve the quality of the results but also do this with much smaller amounts of data.

The key hurdle in a data-driven approach that is often ignored is the difficulty in obtaining adequate quality and quantity of data from a production system. Except in the case of inadvertent mistakes, the configurations that the administrators are willing to use in a production system are extremely limited -- ones that work well. Thus the available data cannot even begin to cover the full range of feasible or even desirable settings. Thus even if we have a huge amount of collected data, its diversity in terms of coverage of the configuration space is extremely limited. Although most production systems do have a small test cluster where any configuration settings are possible,  translating either the configuration settings or the results from the test-system to the production system (or vice versa) is often either infeasible or involves guesswork (and hence significant errors in the data obtained). Thus the basic requirement of a purely data-driven approach, namely, the ability to generate correct and diverse data covering significant portions of the state space, is usually not met in practice. Unfortunately, the current enthusiasm for applying AI/ML techniques often overshadows these considerations. 

Even if arbitrary data gathering is possible in theory, the effort and time required to cover the configuration space make diverse data generation very difficult, as we experienced in our effort to generate CSG data.  {\em This is the main motivation for our performance CHI to be expressed as a scoring system, rather than an exact performance characterization. It is also the motivation to exploit domain knowledge and use experimental data sparingly rather than following a purely data-driven approach which generally requires extensive amounts of data.} Being a coarse-granularity scoring system, the performance CHI is concerned with distinguishing, say, a well-performing configuration from a poor one, as opposed to attempting to do a precise estimation of all relevant CVs for near-optimal performance. Nevertheless, for convenience, we view the performance CHI as a continuous function of the parameter values and evaluate it using both the available data and the domain knowledge. This allows us to substantially reduce the data requirements and yet obtain much better results than a pure data-driven approach. 

The key issue then is how can the domain knowledge be expressed and exploited? It clear that the input provided by the experts must remain rather small even for large problems. Also, we should not expect experts to provide numbers (e.g., the "weights" as in CHeSS) since people tend to make mistakes in providing numbers, and the numbers provided may depend on extraneous factors such as the mood of the person. Instead, we should largely expect experts to provide their insights regarding the system. These insights can often be summarized in the following types of questions:
\begin{enumerate}
    \item Based on the knowledge about the system, which CVs are likely to be at least moderately important for deciding the system performance?
    \item Are certain CVs related by experience based rules of thumb, either precise ones (e.g., each web-server talking to the database needs 10  more DB threads) or fuzzy ones (e.g., each CPU core would add 100-120 MB/s in disk IO requirements)?
    \item Are certain CVs restricted to a certain small set of values (e.g., memory of 32GB, 48GB and 64GB only)?
    \item If the performance generally increases with respect to a CV (e.g., throughput vs. hardware resource amount), is it likely to show a slow decline beyond some point due to increasing overhead (we are not asking the expert what that point is)?
\end{enumerate}
The list above is not intended to be comprehensive but will be used in this paper. A similar approach can be used with additional insights. Also, note that there is no requirement that the domain expert identifies all of these things, but obviously, more information is better as it can reduce the need for data or provide more robust results with the same amount of data. We will also make use of the general principle of "diminishing returns" to ensure a physically plausible and smooth behavior.

One concern that always comes up with respect to human involvement is what if the provided insights are incorrect? This can be addressed to some extent by performing sanity checks based on the available data; for example, if we have a decent amount of data, we could do the principal component analysis (PCA) to determine if the importance provided by PCA generally jives with the one provided by the expert. However, we should not lose sight of the fact that a pure data-driven approach is no panacea, and itself comes with many hazards such as spurious relationships, variations that are opposite to the expected variations, elimination of important variables, overfitting, etc. We demonstrate in this paper that by using the domain knowledge along with the data, we can get more robust results and avoid some of the pitfalls of the pure data-driven approaches. 

The key area of our research is to understand the influence of various CVs ($P$'s) on the observed metric ($O$'s). As CVs can span a wide-dimensional space, a detailed modeling of over 100s of CVs either using a mathematical, simulation, or other technique  is laborious (if not impossible)~\cite{siegmund2015performance,nair2017using,Makrani2021,Masanet2020,Zhou-CM-survey}. Further, such one-off models would suffer from robustness and over-fit, i.e. we need to re-do the model for any change in the configuration space. The design goal of CHI is to produce a \textit{scoring system}, that can give an insight into the configuration space.
That is, CHI should: (i) \textit{discover}  how a CV influences the behavior (i.e. outcome), (ii) give the rate of increase of such an influence, (iii) show the cut-off point for diminishing returns (if any), and (iv) show the rate of decay beyond the cut-off point. Instead of building a detailed performance model, the objective is to discover the influence of various CVs on the observable outcome. 

With these goals, we formulate the following research questions.
\begin{enumerate}[start=1,label={(\bfseries R\arabic*)}]
    %\item \label{r:r1} Understand the influence (or contribution) of individual configuration objects $\{P:p\}$ on the health index $h$.
    \item \label{r:r2} Discover the influence of the CV $P$'s on the health index $h$'s, including the rate of influence, the point of diminishing return or  cut-off point (if any), and the rate of decay (beyond the cut-off point).
    \item \label{r:r3} Correlate the \textit{H} (computed from various $h$'s) of a configuration file to the observed operational metric $O$ (e.g. performance in our case).
    \item \label{r:r4} Determine the $H^{new}$ of a new (unseen data) configuration file such that the new $H^{new}$ should reflect the ``expected behavior'' $O^{new}$ of the new configuration file.
\end{enumerate}

\section{Solution Design}
\label{s:solution}

The purpose of this research is to get an \textit{apriori} metric to express the health of the configuration file, and such a health metric (\textit{H}) should relate to the observable metric ($O$) in the deployed environment.  
We discover the individual health index ($h_{m}$'s) of the configuration  objects $P_m$'s to minimize the error between computed \textit{H}'s and observed metric $O$'s (Eq.~\ref{e:mse}) and thereby determine the $\eta$'s and $\gamma$'s as defined in Eq.~\ref{e:eqMonotonic}  and Eq.~\ref{e:eqUnimodal}. An optimal solution should minimize MSE (ideally zero), thereby relating the health index \textit{H}'s as close as possible to the observed metric $O$'s

\begin{figure}[htb]
\vspace{-3pt}
\centering
\includegraphics[width=0.67\columnwidth]{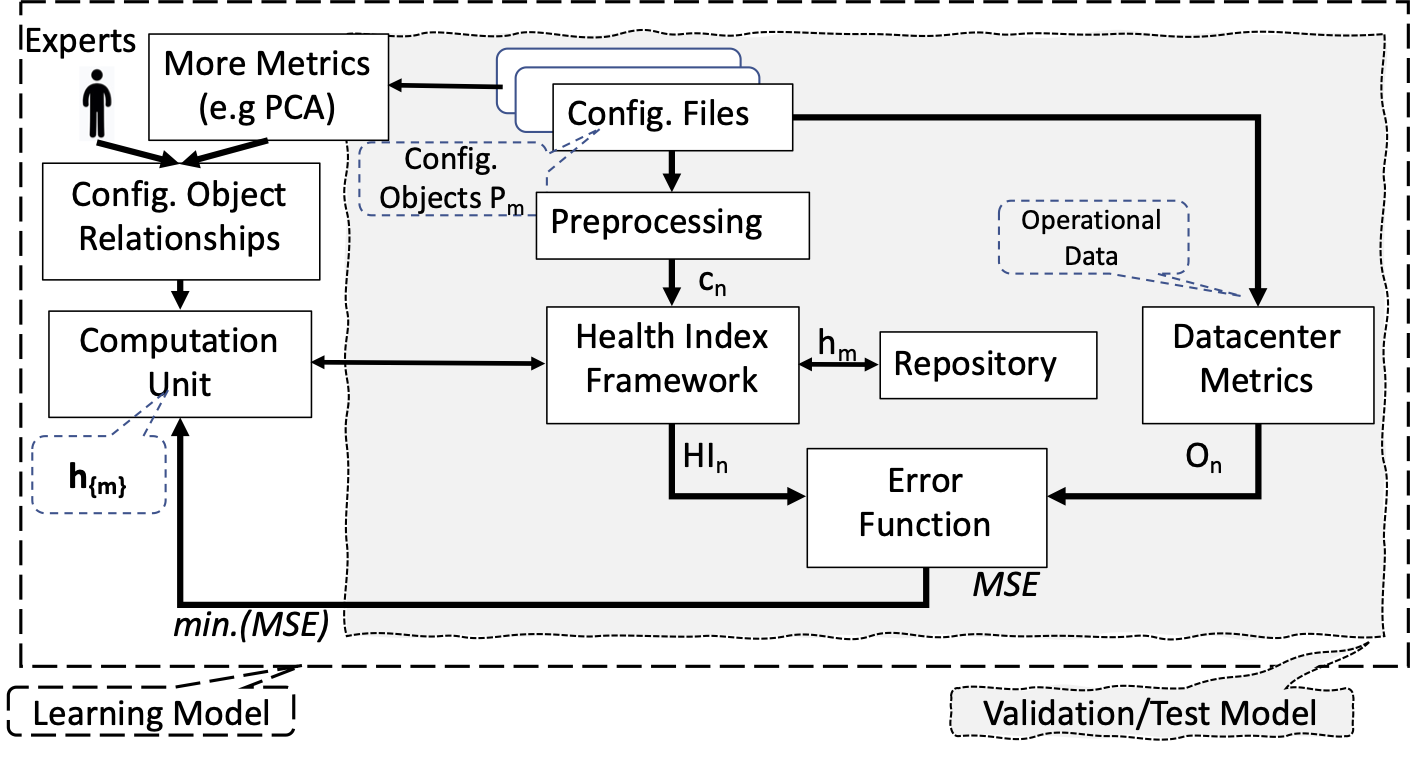}
\vspace{-6pt}
\caption{CHI Framework}
\label{f:framework}
\vspace{-12pt}
\end{figure}

\subsection{CHI Framework}
\label{s:chiframework}

The CHI framework to discover the health index ($H_{n}$'s) of the configuration files ($c_{n}$'s) is shown in  Fig.~\ref{f:framework}. The configuration objects ($P_{m}$'s) in the configuration file is first pre-processed and normalized.

\begin{figure}[htb]
\includegraphics[width=0.75\linewidth]{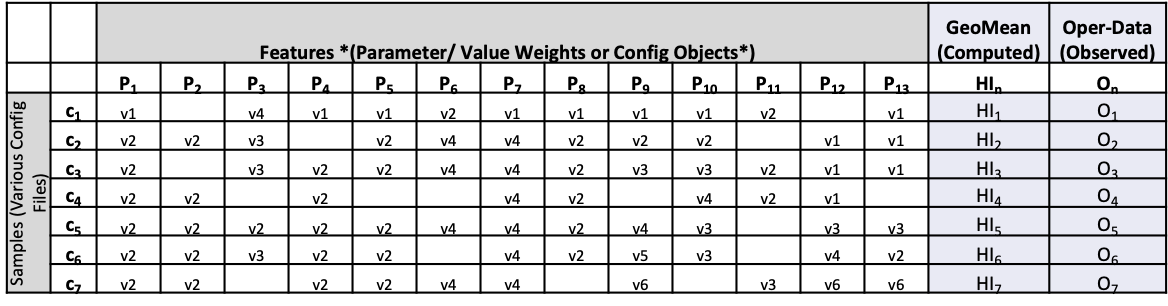}
\vspace{-12pt}
\caption{Abstract Representation of Configuration Files\label{f:abstractData}}
\end{figure}

\begin{figure}[htb]
\includegraphics[width=0.9\linewidth]{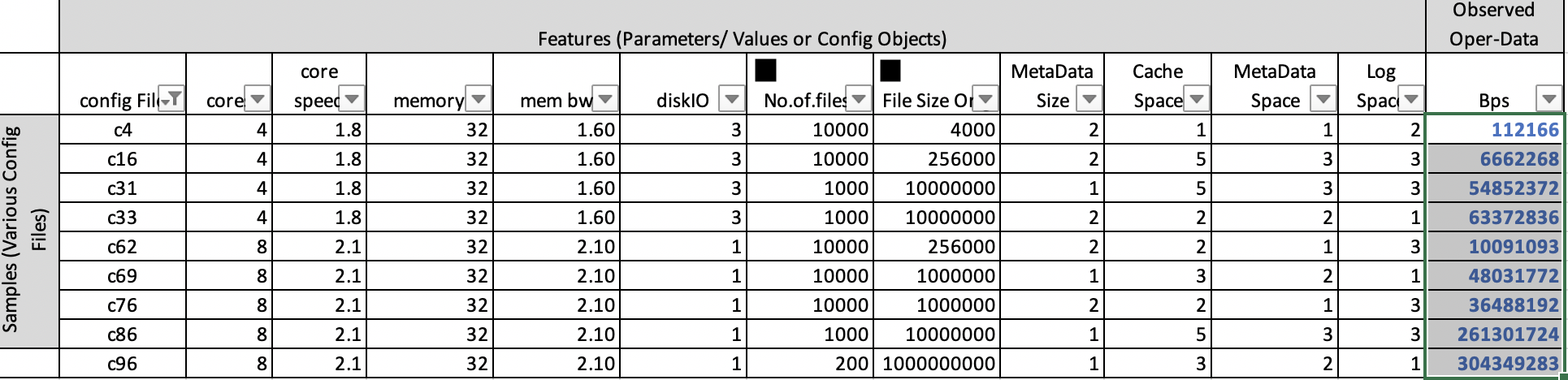}
\vspace{-12pt}
\caption{Sample Configuration File \label{f:sampleData-csg}}
\end{figure}

\begin{figure}[htb]
\vspace{-12pt}
\includegraphics[width=0.9\linewidth]{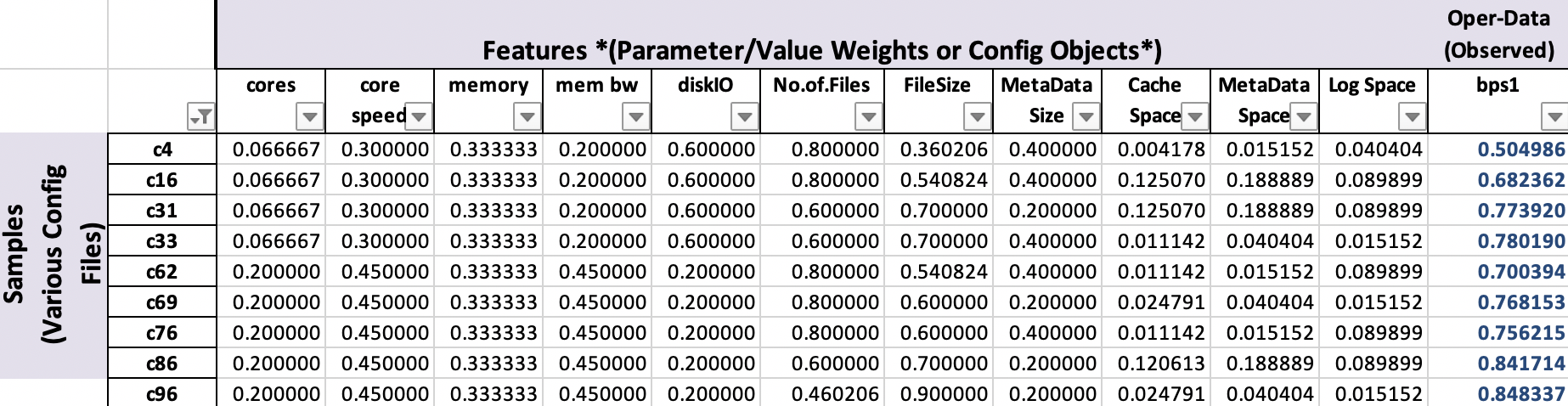}
\caption{Sample Configuration File (normalized)\label{f:sampleDataNorm-csg}}
\end{figure}

Fig.~\ref{f:abstractData} shows an abstract representation of the sample configuration data, with rows illustrating the various configuration files $c_n$, and columns showing the CVs of the configuration $P_m$ with the respective observed metric $O_n$. Each cell $p_{nm}$ represents a name/value pair for the configuration files. Fig.~\ref{f:sampleData-csg} shows a real-world example of the configuration file with the associated observed metric (i.e. performance in Bps shown in the last column as $O_i$). The normalized values of the configuration objects and corresponding normalized operational metrics form the basic input to the framework. A sample of a normalized version of the input files is illustrated in Fig.~\ref{f:sampleDataNorm-csg}. Domain experts or service specifications define the boundaries of the configuration object, i.e $P^{(min)}$ \& $P^{(max)}$. As part of pre-processing and to keep the format of all input data uniform, it may sometimes be necessary to fill in any undefined/missing values (shown as blank cells in Fig.~\ref{f:abstractData}). If necessary  feature engineering methods have to be incorporated to enrich or supplement an existing feature (i.e configuration object $P_m$) with a new feature (i.e configuration object $P'_m$).

\begin{table}[htb]
\begin{minipage}{0.45\textwidth}
\footnotesize
\begin{tabular}{|P{0.5in}|p{1.80in}|}
    \hline
    $M$ & Number of CVs \hh 
    $N$ & Number of configuration files \hh
    $P_m$ & $m^{th}$ CV in a configuration file ($0 \le m \le M$)\hh
    $c_n$ & $n^{th}$ configuration file ($0 \le n \le N$) \hh
    $p_{nm}$ & name/value pair $m$ of configuration file $n$ \hh
    $h_{nm}$ & health index (aka weight) of $p_{nm}$ \hh
    $H_n$ & Health Index of $n^{th}$ configuration file  \hh
    $O_n$ & Observed Metric of $n^{th}$ configuration file\hh
    $L_{sd}$ & Strong dependent CVs \hh
    $L_{wd}$ & Weakly dependent CVs \hh
    $L_{un}$ & Unimportant ones \hh
    $L$ & $L=M-L_{sd}-L_{wd}-L_{un}$ dominant CVs \hh
    $f_{mk}()$ & Relationship function between CVs $p_{m}$ \& $p_{k}$ \hh
    $s_{nm}$ & Normalized value of $p_{nm}$ \hh
\end{tabular}
\caption{Nomenclature used in the paper. \label{t:nomenclature}}
\end{minipage}
\qquad
\begin{minipage}{0.5\textwidth}
\vspace{-12pt}
\captionsetup{type=figure} 
    \includegraphics[width=0.82\linewidth]{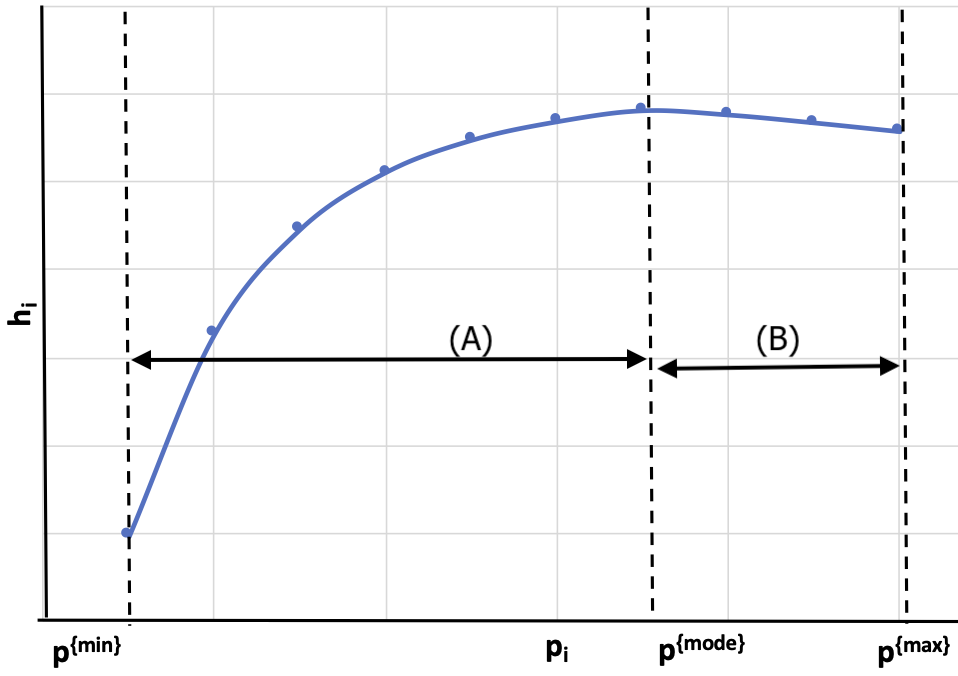}
    \vspace{-12pt}
    \caption{Sample CV Value vs. \textit{Health Index} relationship\label{f:figureP}}
\end{minipage}
\end{table}

\subsection{Estimating Health Index From Configuration Data}
\label{s:estimation}

Our data $\DD$ is a set of {\em distinct} $N$ configuration files, or "rows", say $c_1,..,c_N$ with  configuration $c_n$, $n\in 1..N$ and its corresponding observed output  $O_n$ (e.g., performance) (See sample file in Fig.~\ref{f:abstractData}). A configuration is defined by a set of $M$ CVs (or "columns"), denoted $P_1,..,P_M$. That is, each configuration $c_n$ is a vector of $M$ `values' for CVs $P_1,..,P_M$, henceforth denoted as $p_{n1},..,p_{nM}$. We postulate the health index $H_n$ for each $c_n$, which itself is computed as a geometric mean of \textit{H} of individual CVs (Eq.~\ref{e:HI}). The \textit{H} for an individual CV is denoted as $h_{nm}$, $m=1..M$, for each CV value $p_{nm}$. Our goal is to estimate $h_{nm}$'s, and hence $H_n$s, compatible with the observed outputs $O_n$. {We pose this as an optimization problem.} The assumptions and constraints are as follows.

\begin{enumerate}[start=1,label={(\bfseries I\arabic*)}] 

\item \label{i:item1} Of the $M$ CVs, $L_{sd}<M$ CVs may be {\em strongly dependent} on others, and we assume that this relationship, denoted as $\models$.
%, is \KKC{given by the experts}. 
Thus, if parm $P_m\models P_k$, then $p_{nk}$ is functionally determined by $p_{nm}$ for all $n$, i.e., $p_{nk}=f_{mk}(p_{nm})$ where $f_{mk}$ is a known analytic function that transforms column $m$ to column $k$ (independent of the row index $n$).

\item \label{i:item2} In addition, another $L_{wd}<M$ CVs may be {\em weakly dependent} on others, and we also assume that this relationship, denoted as $\mapsto$, is given by the experts. Thus, if parm $P_n\mapsto P_k$, then the value $p_{nm}$ restricts the choice of values for $p_{nk}$ to a small range around some value, i.e.,

\EQ
p_{nk}=f_{mk}(p_{nm}) (1+r_k), \qs r_k\in [-R_k..R_k], k\in\LL_{wd} 
\label{e:Lwd}
\EN
\newline
where $r_k$ represents the uncertainty as a fraction. Here $R_k\in 0..1$ is a small (known) fractional number representing the boundaries of the uncertainty. For example, $R_k=0.1$ means that the value of $p_{nk}$ can vary $\pm 10\%$ around the value determined by the function $f_{mk}$.

\item \label{i:item3} The relatively important CVs are usually known to the experts from experience or could be obtained using a statistical technique like principle component analysis (PCA). Others are better  eliminated since marginally important CVs only tend to increase the noise in the estimations~\cite{mass2019sondur}. We assume that of the $M-L_{sd}-L_{wd}$ primary CVs, $L_{un}$ are unimportant and hence eliminated. Thus, we are left with only $L=M-L_{sd}-L_{wd}-L_{un}$ CVs.  The normal $L_{\cdots}$ parameters introduced above represent the sizes of the sets  $\LL_{\cdots}$ (i.e. $L_{\cdots}$ =  $|\LL_{\cdots}|$).

\item \label{i:item4} Based on the last few points, we only need to consider $L$ CVs in the formulation. For convenience, we denote the corresponding {\em set} of variables of different types as $\LL_{..}$, i.e., $\LL_{sd}$ is set of strongly dependent CVs, $\LL_{wd}$ is set of weakly dependent variables, etc.

\item \label{i:item5} We postulate two different forms of functions $h_{nm}$s that we want to estimate -- monotonic and unimodal. Of the $L$ CVs, we assume that $L_{mo}$ is monotonic and $L_{um}$ is unimodal. As before, we represent the corresponding CV sets as $\LL_{mo}$ and $\LL_{um}$ respectively. This behavior is illustrated as an example in Fig.~\ref{f:figureP}, with monotonic behavior depicted in the area (A) and unimodal behavior beyond point $p^{(mode)}$ in the area (B).

As explained earlier, we assume that the minimum and maximum values of the CV, denoted $p_m^{(min)}$ and $p_m^{(max)}$ respectively are defined by experts. We define $p_m^{(min)}$ as the value for which $h_{nm}=h^{(min)}_{nm}$. 
%It is crucial to realize that by this definition, no working configuration will ever choose $p_{nm}=p_m^{(min)}$, since this would correspond to zero performance. In fact, all practical configurations will have $p_m$ closer to $p_m^{(max)}$.  
Now we have two cases:
%\begin{enumerate}

{\bf Monotonic:} $h_{nm}$ increases monotonically with $p_{nm}$ for CV $m$ and when $p_{nm}=p_m^{(max)}$, $h_{nm}=h^{(max)}_{nm}$. We expect the relationship to be concave (i.e., follow law of diminishing returns). We capture this using the equation:

\EQa
s_{nm} \aeq \frac{p_{nm}-p_m^{(min)}}{p_m^{(max)}-p_m^{(min)}} \\
h_{nm} \aeq \frac{1 - e^{-\eta_m s_{nm}}}{1-e^{-\eta_m}},\hqs m\in\LL_{um}, s_{nm},\hqs m\in\LL_{mo} \label{e:eqMonotonic}
\ENa

where $\eta_m$ is a predefined positive parameter that controls the growth rate. 

{\bf Unimodal:} Here we assume the same equation as above, except that the maximum happens at the value $p_m^{(mode)}<p_m^{(max)}$. Beyond $p_m^{(mode)}$, we can assume that $h_{nm}$ decreases linearly with maximum fractional degradation of $\gamma_m<1$. (Generally, $\gamma_m\ll 1$) 

\EQa
h_{nm} \aeq \frac{1 - e^{-\eta_m s_{nm}}}{1-e^{-\eta_m}},\hqs m\in\LL_{um}, s_{nm}\le p_m^{(mode)}  \\
h_{nm} \aeq 1 - \gamma_m s_{mn},\hqs m\in\LL_{um}, s_{nm} > p_m^{(mode)} \label{e:eqUnimodal}
\ENa
where
\EQ
s_{nm} = \frac{p_{nm}- p_m^{(mode)}}{p_m^{(max)}-p_m^{(mode)}} \nn
\newline
\EN
\vspace{6pt}
\end{enumerate}

Fig.~\ref{f:figureP} represents the relationship represented in Eq.~\ref{e:eqMonotonic},  and Eq.~\ref{e:eqUnimodal} and depicts an example behavior of a CV. $h_{nm}$ for the CV will increase monotonically up to a limit $p^{(mode)}$, and then linearly decreases beyond $p^{(mode)}$. We now have to predict the $h_{nm}$ contribution of the CV $p_{m}$ given these boundaries. The total number of unknowns is thus $2L+L_{wd}+L_{um}$, and we expect that the number of rows $N$ (i.e., configurations for which output is known) will be significantly larger than the $M$.

\textbf{Objective:} The objective now is to determine the unknowns introduced above, i.e., $r_k, k\in\LL_{wd}$, and $\eta_m$, $m\in\LL_{mo}$ and $\gamma_m$, $m\in\LL_{um}$  to minimize the mean square error (MSE) between the estimated $H_n$'s and observed output $O_n$'s. MSE is given as:  

\EQ
MSE = \frac{1}{N} \sum_{n=1}^{N} (H_n - O_n)^2 
\label{e:mse1}  \nn
\EN

or alternatively,
\EQ
MSE = \frac{1}{N} \sum_{n=1}^{N} (log(H_n) - log(O_n))^2 
\label{e:mse} 
\EN

\subsection{Computing CHI}

With inputs about the CV boundaries ($P^{(min)}$ \& $P^{(max)}$) and the pre-processed normalized configuration files, the CHI framework  computes the health index $h_{nm}$ using a  non-linear gradient descent regression  model to achieve the desired objective (i.e. minimize the MSE). MSE is hierarchically dependent on  other variables as explained in section~\ref{s:estimation}  (item~\ref{i:item1} to item ~\ref{i:item5}). The gradient of MSE ($\nabla MSE$) w.r.t individual dependent variable $\kappa_{nm}$ is represented in Eq.~\ref{e:gradmse}, and split into three components: (i) MSE is a function of \textit{H} (hence, $\partial MSE/\partial H$), (ii) \textit{H} in turn, depends on individual $h_{nm}$ (hence the second part: $\partial H_n/\partial h_{nm}$), and (iii) individual $h_{nm}$ is a function of either $\eta_m$ or $\gamma_m$ (hence the final derivative). To minimize MSE, we employ a gradient descent algorithm, with each iteration calculating a new state $\kappa$  computed as a function of $\alpha$ \& $\nabla MSE$ as shown in Eq.~\ref{e:ridgereg} (where $\alpha$ represents the learning rate).

\EQa
\nabla MSE \aeq \frac{\partial MSE}{\partial \kappa_{nm}} = \frac{\partial MSE}{\partial H_n} * \frac{\partial H_n}{\partial h_{nm}} * \Psi_{nm} 
\label{e:gradmse} 
\ENa

\EQa
\text{where }
\Psi_{nm} = {
\begin{cases}
    \frac{\partial h_{nm}}{\partial \eta_{nm}},   & \text{if } s_{nm} \le p_m^{(mode)} \\
     \frac{\partial h_{nm}}{\partial \gamma_{nm}}  & \text{otherwise}
\end{cases}}
\ENa

\EQa
\kappa_{nm} \leftarrow \kappa_{nm} - \alpha \nabla MSE 
\label{e:ridgereg} 
\ENa

\begin{figure}[htb]
\vspace{-12pt}
\centering
\begin{tabular}{cc}
\subfigure[{Compute Unit}]{
\includegraphics[height=1.5in,width=0.45\columnwidth]{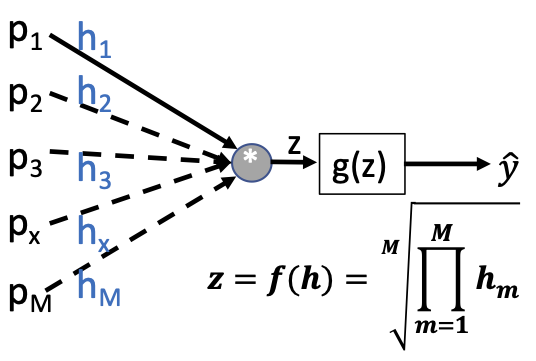}
\label{f:computeNode}} &
%\vspace{-6pt}
\subfigure[{Neural Network ({NN})}]{
\includegraphics[height=1.5in,width=0.45\columnwidth]{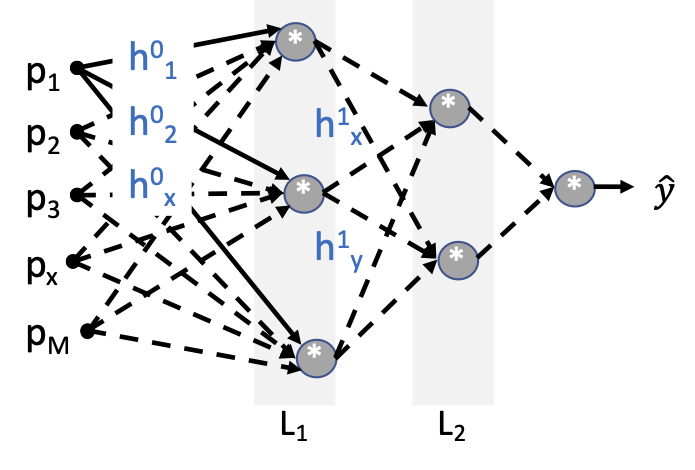}
\label{f:neuralNetwork}}
\end{tabular}
%\vspace{-6pt}
\caption{CHI Compute Unit Design}
\label{f:figureP1}
\vspace{-12pt}
\end{figure}

\EQa
\text{where }
\kappa_{nm} = {
\begin{cases}
    \eta_{nm}  & \text{if } s_{nm} \le p_m^{(mode)}\\
     \gamma_{nm}  & \text{otherwise}
\end{cases}}
\ENa
\newline
The CHI computation unit in Fig.~\ref{f:computeNode} represents the calculation of the \textit{H} (Eq.~\ref{e:HI}). The unit regresses and computes $H$, $\nabla MSE$, \& $\kappa_{nm}$ (as given above).  An error function computes the difference between ``computed'' $H_n$ and observed operational metric $O_n$, and updates the $\nabla MSE$ \& $\kappa_{nm}$ for the next states (Eq.~\ref{e:ridgereg}). The algorithm terminates after it reaches a predefined termination condition (either expressed as the number of iterations or on achieving the desired MSE).   At  termination, the algorithm persists the `discovered' relationships $\eta_{m}$'s \& $\gamma_{m}$'s of the  configuration object $P_{m}$'s in the local repository, so that these relationships can be referred in the future to compute the $H^{new}$ of a ``new/unseen'' configuration files. In the results section, we show the effectiveness of the algorithm  in discovering the influence of $P_m$'s on $O_n$'s and the \textit{computational  accuracy of health indices} (i.e. $h_{m}$'s). We show that CHI can discover the unknown's given above to satisfy the objective (minimize MSE) and that they relate the contribution of various CVs $P:p$ to the health index $h$ of the configuration object (and indirectly to the $O$'s).

\subsection{CHI Compute Unit Design}
\label{s:computeunitdesign}
 
The design  in Fig.~\ref{f:computeNode} represents the \textit{H} computation in Eq.~\ref{e:HI}, with $p$'s representing the configuration object values ($p_{nm}$) and the weights ($h$'s) are the contributions of the $p_{nm}$ on the health index $h_{nm}$ as defined by  Eq.~\ref{e:eqMonotonic},  and Eq.~\ref{e:eqUnimodal}. In the traditional ML, a neuron computes an estimated output value $\hat{y}$  equal to the weighed ($w$) sum of the input features ($x$), i.e. $\hat{y}=1/M\sum_{m=1}^{M} (w_m.x_m)$. Following similar concepts, we design  the CHI computation unit to represent the geometric mean \textit{H} of the configuration objects in a configuration file, i.e. $\hat{y}=\sqrt[{M}]{\left(\prod _{m=1}^{M} f(x_m)\right)}$, where $f(x_m)$ represents the health index function the individual CV ($c_m)$ on the outcome. Thus, $h_m$=$f(x_m)$ is the unknown and needs to be learned. 

CHI compute unit \textit{acts solely} on the weights ($h_m$=$f(x_m)$) which in turn is a function of $\eta_m$'s \& $\lambda_m$'s. With a  CHI design as above, the required solution is the estimation of the parameters $\eta_m$ \& $\gamma_m$ that contribute to the (weight) health index $h$'s such that it minimizes the mean square error (Eq.~\ref{e:mse}).  The transfer function $g(z)$ represents the non-linearity in the model, represented a: $ g(z) = max(\epsilon,z)$ where $\epsilon$ is a small value ($10^{-3})$ to ensure a small positive gradient. This transfer function~\cite{daubechies2019nonlinear}~\footnote{Referred in the ML literature as Leaky Rectified Linear Unit (Leaky ReLU)} ensures that $h_{m}$ is always positive and allows the complex relationships in the data to be learned.

\subsubsection{CHI is not a neural network model:} In the ML domain, a neuron network ($\mathbb{NN}$) algorithm is built using the neuron as a basic unit, and is used to solve a non-linear problem.  Fig.~\ref{f:neuralNetwork} illustrates such an example with two hidden layers (shaded bands) and three neurons in layer $L_1$ and two neurons in layer $L_2$. In practice, $\mathbb{NN}$ can have a large number of such neurons and hidden layers. Though our approach is a non-linear regression model, we argue that CHI  \underline{cannot} be represented as a  $\mathbb{NN}$ because: (i) while the $\mathbb{NN}$ can predict a new $H^{new}$ with a prior training set of configuration files $C_{n}$'s, the $\mathbb{NN}$ model cannot discover the influencing factors of the individual $P_{m}$'s \textit{itself}, (ii) if  $h^k_l$ represents the $l^{th}$ neuron in $k^{th}$ layer (for some $l,k$) in Fig.~\ref{f:neuralNetwork}, then combining all the individual (weights) health indices from all the neurons to represent the health index ($h_{m}$) of input configuration object is unnecessarily complex, (iii) incorporating \textit{H} relationships as given in  Eq.~\ref{e:eqMonotonic},  and Eq.~\ref{e:eqUnimodal} is difficult, and (iv) importantly, $\mathbb{NN}$  does not give a monotonic or unimodal relationship between object $P_{m}$ and its influence (rate of influence, cut-off point, rate of decay, etc.) as discussed in the research goal.

Although the  non-linear gradient descent regression  model can be further improved with robust loss function and optimization techniques, our approach did not venture into  this  area.

\subsection{Identifying Unimportant CVs}
\label{s:unwantedconfig}

It is well known that the configuration space is too huge to explore and to get a data set covering all known combinations of configuration objects. It has been observed that the software performance functions are usually very sparse i.e. only a small number of configurations and their interactions have a significant impact on system performance~\cite{ha2019deepperf}. Various tools and techniques are being explored to limit such configuration spaces~\cite{krishna2020conex,pereira2019learning}.  Most literature  agree that domain expertise is often the best and fastest way to eliminate unwanted features (configuration objects in the problem)~\cite{krishna2020conex}. Instead of relying on a pure human approach or trusting a generic algorithm to sort the important and unimportant CVs, our approach eliminates the unimportant CVs ($L_{un}$) with Principal Component Analysis  (PCA) ``assisted'' domain expertise. PCA is a dimensionality reduction technique that projects the data from its original $p$-dimensional space to a smaller $k$-dimensional subspace. By using PCA, domain experts can confirm their belief on which CVs are of importance versus the unimportant CVs ($L_{un}$). For example, using PCA software-analytics researchers recursively divide data into smaller or as a preprocessor tool  to reduce noise in software-related data sets~\cite{theisen2015approximating, nair2018faster,Cloud2019Sondur}.

\section{Experiments and Empirical Data}
\label{s:experiments}

In this section, we present the data-set used and its characteristics followed by a detailed evaluation of results. All code was developed in Python and all evaluations were run on a MacBook Pro 2.5 GHz x 2 core Intel i7 with 16 GB memory.

\subsection{Data Sets and Hyperparameters}
\label{s:data sets}
A detailed study of system configuration and performance needs a well-defined data-set that captures the resource allocation (i.e. configuration settings) and observed behavior (e.g. performance) under various conditions (e.g. hardware servers, workload, etc.). There are many publicly available data sets as described by Google~\cite{reiss2011google}, Alibaba~\cite{guo2019limits},  and other Cloud traces~\cite{otherTraces} capture large time-series data for measures such as CPU utilization, IO rates, network traffic, etc.; unfortunately, they are not useful for us since configuration information is invariably missing. In fact, for some of the data-sets, the configuration continues to change dynamically, but there is no information about it. 

We did locate some real world configuration data sets in~\cite{nair2017using,TUDelftBitBrains} but there is still some question about the configuration settings, especially for~\cite{TUDelftBitBrains}. We also use the CSG dataset that we have created ourselves~\cite{Cloud2019Sondur}. It has several advantages over others including most of all a complete control over and knowledge of the configurations used, data collected, and difficulties encountered. In particular, not surprisingly, we found that about 2-3\% of the cases did not produce sensible results. These experiments had to be repeated and in about 1\% of the cases the data discarded due to unexpected interference during the experiment. This makes the CSG data much cleaner and usable than others. We obtained the data with some variations in both the hardware setup and the workloads. Note that the hardware setting variations are generally missing from other data sets.  However, due to the substantial effort and time required in setting up a different configuration and conducting each experiment (about 1-2 hours), we obtained only about 1000 measurements, of which 990 were retained.  
Before outlining the evaluation (and for completeness), we briefly introduce the real world data-set collected from a Cloud Storage Gateway based on our earlier work in section
~\ref{s:data-csg}.

The usable public data sets that we found for our configuration studies are listed in Table~\ref{t:dataset}. We ran our CHI framework on all of these to answer the research questions (\ref{r:r2} to \ref{r:r4}) discussed above. We followed established  practice similar to an ML approach:  the complete data  $\DD$  ($c_n$'s \& $O_n$'s) is first normalized and randomly split into two groups - train ($\DD_{train}$) and test ($\DD_{test}$). We evaluate using two cases: (i) 50\% $\DD_{train}$ \& 50\% $\DD_{test}$ and (ii) 80\% $\DD_{train}$ \& 20\% $\DD_{test}$. The $\DD_{train}$ is input to the CHI model to compute and discover the unknowns $\gamma_m$'s, $\eta_m$'s \& $h_{nm}$'s of various $P_{m}$'s using the steps explained above. 

To maintain uniformity across all studies and test cases, we maintained the iteration limit (i.e. epochs) to 500 and learning rate $\alpha$ to 0.5 and observed that the CHI reaches a satisfactory MSE  (i.e min $\nabla$ MSE) during these epochs, and there is no significant improvement afterwards. The resulting `learnt' values of $\gamma$'s \& $\eta$'s of various $P_m$'s (from the training data $\DD_{train}$) is stored in a repository and used to calculate the new health index $H^{test}_i$ of the {\em unseen test configuration} from $\DD_{test}$. We compute the error rate as the difference between computed health index $H^{test}_i$ representing the ``expected performance'' and observed performance ($O$'s) (as given in Eq.~\ref{e:mse1}). The error rate (MSE and variance) of the newly predicted health index ($H_i$'s vs. $O_i$'s) is given in Table ~\ref{t:dataset} for the two test-cases. Our focus is on understanding the influence of CVs, rather than a performance prediction model, hence we did not venture into detailed ML evaluation metrics such as k-Fold evaluation\footnote{Though the above 80/20 test results can represent one of the k-Fold results (for k=5).}, recall, precision, etc.

\begin{table*}[htb]
\begin{minipage}{\textwidth}
%\vspace{-12pt}
\caption{Data-set used in the paper.}
\centering
\begin{tabular}{|p{0.25in}|p{1.4in}|p{1.00in}|P{0.25in}|P{0.35in}|P{0.30in}|P{0.4in}|P{0.30in}|P{0.4in}|} \hline
\multirow{2}{*}{Code}  & \multirow{2}{*}{System [Related Art]} & \multirow{2}{*}{Domain} & \#Attrs & Samples & \multicolumn{2}{c|}{A (50/50)} & \multicolumn{2}{c|}{B (80/20)} \\
   &  &  &  (M) & (N) & MSE & Variance & MSE & Variance \\
   \hline
 CSG\footnote{[CSG] \url{https://www.kkant.net/config_traces/CHIproject}\hfill\:}
 & Cloud Storage Gateway~\cite{Cloud2019Sondur} & Cloud Storage & 10 & 105 & 0.0121 & 0.0068 & 0.0098 & 0.0058 \hh
 BB\footnote{[BB] \url{http://gwa.ewi.tudelft.nl/datasets/gwa-t-12-bitbrains} (RND500)\hfill\:} & BitBrains Datacenter~\cite{shen2015statistical} & Virtual Machines & 7 & 390 & 0.0526 & 0.0256 & 0.0475 & 0.0236 \hh
 SS2\footnote{[SS2,SS3,SS8,SS10] \url{https://github.com/ai-se/ActiveConfig_codebase/tree/master/RawData}\hfill\:} & SQL Lite~\cite{nair2018finding} & SQL server & 29 & 2000 & 0.0620 & 0.0311 & 0.0583 & 0.0308 \hh
 SS3 & Berkeley DB C~\cite{nair2018faster} & Embedded database & 18 & 2000 & 0.0417 & 0.0219 & 0.0332 & 0.0177 \hh
 SS8 & Apache~\cite{siegmund2015performance} & Web Server & 9 & 2000 & 0.0371 & 0.0212 & 0.0316 & 0.0167 \hh
 SS10\footnote{[SS2,SS3,SS8,SS10] \url{https://goo.gl/689Dve} (RawData/PopulationArchives)\hfill\:} & Roll Sort~\cite{nair2018faster} & Sorting Tool & 6 & 3840 & 0.1887 & 0.0944 & 0.1842 & 0.0932 \hh
\end{tabular}
\label{t:dataset}
%\vspace{-18pt}
\end{minipage}
\end{table*}

\subsection{Data-set Characteristics}
\label{s:data-characteristics}

\subsubsection{Cloud Storage Gateway (CSG) Data-set}
\label{s:data-csg}
\begin{figure}[htb]
\vspace{-6pt}
\centering
\includegraphics[width=0.6\columnwidth]{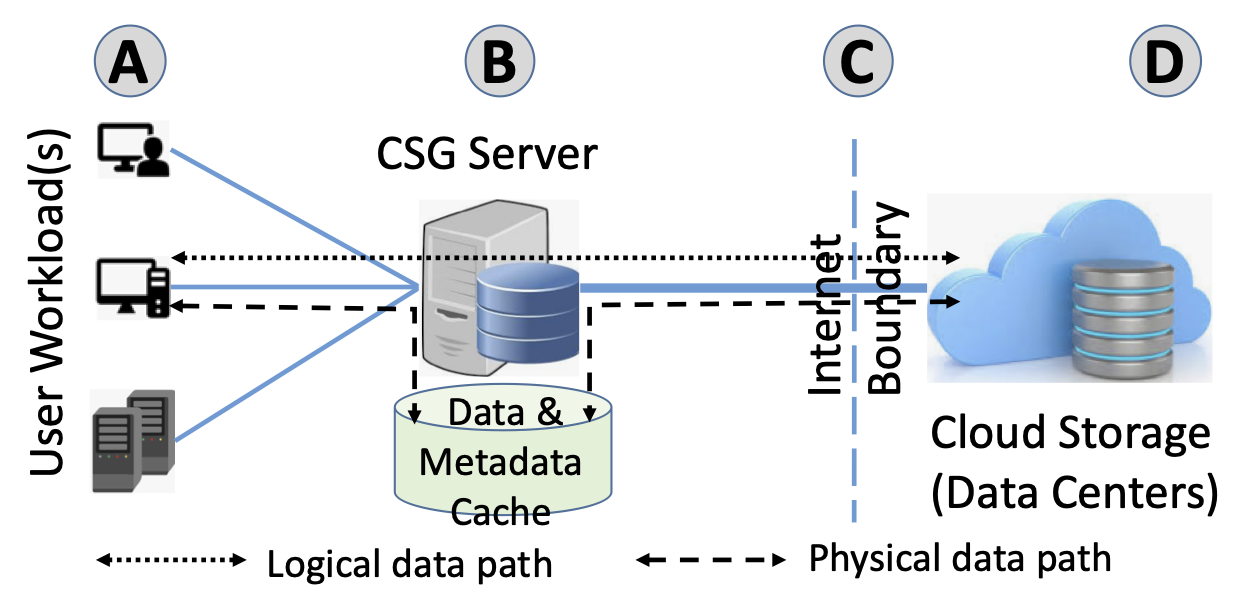}
\vspace{-6pt}
\caption{Edge Computing/ Cloud Storage Gateway}
\label{f:CSG}
\vspace{-12pt}
\end{figure}

CSG is architecturally similar to Edge Computing, IoT Gateways, etc. which are constrained by limited resource capacity and placed between the Edge/IoT/user applications and the Cloud platform.
Fig.~\ref{f:CSG} conceptually shows the CSG operation. A CSG is usually deployed at a branch office or remote location and has access to a rather limited local compute/storage and is connected to a Cloud data center over the Internet. A CSG  essentially uses local storage as a cache for the remote Cloud storage to  bridge  the gap  between  the demand for low-latency/high-throughput local access and the reality of high-latency connection to the cloud with unpredictable and usually low throughput. 

The observed performance of CSG denoted as $O$, is influenced by its configuration variables (CVs), denoted as $P_i$ for $i$th CV. The CVs include compute resources (cores, cpu-speed, memory capacity etc.), IO path (memory bandwidth, disk IO bandwidth, etc), buffer space allocation (cache space, meta-data space), etc.  A full description of the CSG system, various CVs influencing the behavior, real-world experiments to collect empirical data is given in our earlier paper~\cite{Cloud2019Sondur}.  We ran about 1000 experiments and collected data on different configurations (denoted $c_n, n=1,2,..,N$). Each configuration $c_n$ involves the setting of $M$ different configuration variables (CVs).  This data-set was further averaged and smoothed the outliners (to a final data-set of 105). Fig.~\ref{f:sampleData-csg} illustrates an abstract view of the configurations $c_n$'s, the corresponding outputs $O_n$'s (known), and the H's (to be estimated).

Our CSG configurations include CPU cores, DRAM bandwidth, memory capacity, and storage bandwidth during the execution of workloads (inline with Ref.~\cite{Makrani2021}), although the number of variations that we experimented with had to be limited for practical reasons. Nevertheless, the availability of both hardware and software parameters in our data helps us do a good evaluation and to better explain the results below.

The workload is an important component that defines the behavior of the system and the observable outcome (e.g. performance)~\cite{papadopoulos2016peas}. In CSG data-set $\DD$, the number of files, file size, and request metadata size refers to the user workload (provided by the vendor). Applying the principles stated in section~\ref{s:estimation},  we eliminated  the least important CVs (i.e. $L_{un}$ above). For example, using domain knowledge  coupled with PCA and reasons explained in the CSG paper~\cite{Cloud2019Sondur}, we marked Log Space Resource and Network Bandwidth as unwanted CVs ($L_{un}$). The normalized data-set of the empirical data is shown in Fig.~\ref{f:sampleDataNorm-csg}. Based on the widespread of a few data-points (e.g. file size, and no. of files), we used Log normalization to re-engineer the configuration object values ($p_{nm}$) to a new object ($p'_{nm}$). The full data-set was normalized between a small value ($\epsilon = 10^{-3}$) and 1.0. After such pre-processing, we use the data-set in the CHI to discover individual $\gamma$'s \& $\eta$'s of various $P_{m}$'s.  With this empirical data in hand, we  applied the CHI to answer the research questions (\ref{r:r2} to \ref{r:r4}) raised above.

\begin{figure*}[!ht]
\includegraphics[width=0.9\linewidth]{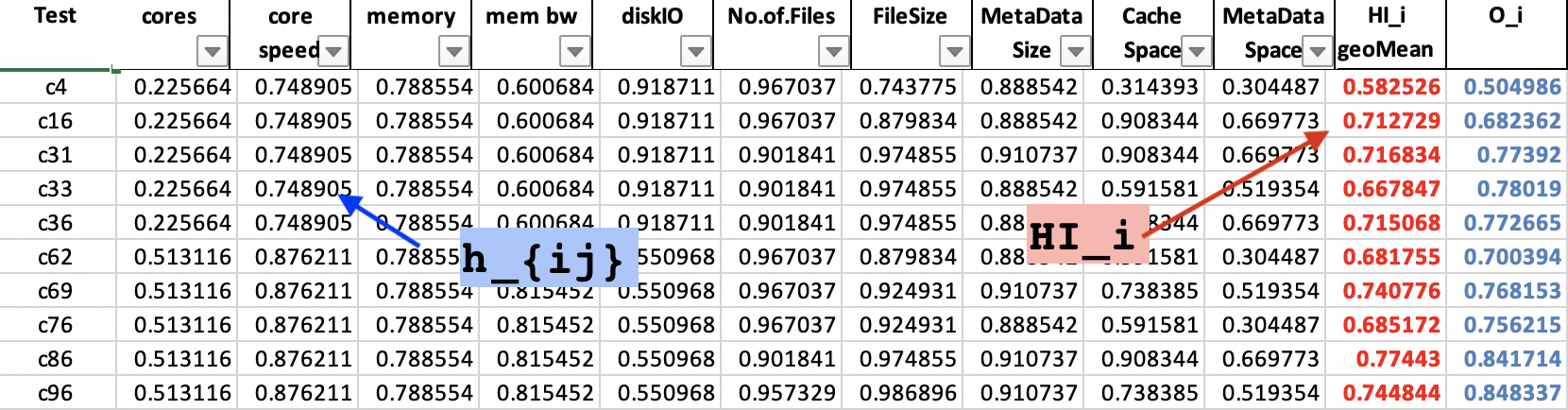}
\caption{Results: Configuration Files with $h$ \& \textit{H} metrics}
\label{f:resultsData-csg}
\vspace{-6pt}
\end{figure*}

\subsubsection{"BitBrains" Data-set}
\label{s:bb-dataset}
Next, we examine the application of CHI to the public domain data from TU Delft BitBrains data-trace~\cite{TUDelftBitBrains}. This data-set contains the performance metrics of 1,750 Virtual Machines (VMs) from a distributed data-center from BitBrains, which provides specialized services for managed hosting and business computation for enterprises. This data-set includes some mixture of customer workload from major banks (e.g., ING), credit card operators (e.g., ICS), insurers (e.g., Aegon), etc. During pre-processing, we noticed that the 'fastStorage-1250' data-set contained huge records of zero values (e.g. zero disk IO or network activity) compared to the'Rnd-500' data-set. Therefore, we used the latter, which has 500 VMs that are either connected to the fast SAN (storage area network) systems or to much slower Network Attached Storage (NAS) systems. The data characteristic and  usage is described in Table.~\ref{t:bbdata-set}.

\subsubsection{"Enterprise" Data-set Characteristics}
\label{s:enterprise-dataset}

The last four data sets (SS2, SS3, SS8, SS10)  in Table.~\ref{t:dataset} have been used in~\cite{siegmund2015performance, nair2017using,nair2018faster} for the performance model, that we compare against our approach. For simplicity, we label these as  "Enterprise Data-set". These data sets include traces from a web-server, key-value DBMS, relational DBMS, and a sorting tool.  
%\KKC{Need to talk about CVs of all these workloads, not just Apache}
Berkeley DB (C) (marked SS2) is an embedded key-value-based database library that provides scalable high performance database management services to applications. SQLite (SS3) is the most popular lightweight relational database management system used by several browsers and operating systems as an embedded database. Apache HTTP Server (SS8) is a highly popular Web Server.
Incidentally, the Apache server has about 550+~\cite{xu2015hey} CVs\footnote{Apache doc. at: \url{https://httpd.apache.org/docs/2.4/configuring.html} \& \url{https://httpd.apache.org/docs/2.4/mod/core.html}} but these were cut-down to only nine CVs in~\cite{siegmund2015performance, nair2017using}, but the rationale or the method for doing so is unclear.  Roll Sort (SS10) is an environment configuration where rs is run by varying 6 features and the throughput is measured. The characteristics of the Enterprise data-set\footnote{Data-set at: \url{https://github.com/ai-se/Reimplement/tree/cleaned_version}} is given in Table.~\ref{t:datasetEnterprise} and  the description\footnote{CV details: \url{http://tiny.cc/3wpwly}} of the CVs is taken from Ref.~\cite{nair2017using}.
We refer readers to the detailed  literature at Ref.~\cite{siegmund2015performance, nair2017using,nair2018faster} for full systems description of these data sets.

\begin{table*}[htb]
%\vspace{-12pt}
\caption{Characteristics of Enterprise Data-set~\cite{nair2018finding,nair2018faster}}
\centering
\begin{tabular}{|p{0.25in}|p{0.80in}|p{3.60in}|P{0.60in}|} \hline
{Code}  & {System} & {Description of CVs} & {Observed Behavior}\hh
 SS2 & SQL Lite server & \textit{OperatingSystemCharacteristics, SQLITESECUREDELETE, ChooseSQLITETEMPSTORE, SQLITETEMPSTOREzero, SQLITETEMPSTOREone, SQLITETEMPSTOREtwo, SQLITETEMPSTOREthree, EnableFeatures, SQLITEENABLEATOMICWRITE, SQLITEENABLESTAT2, DisableFeatures, SQLITEDISABLELFS, SQLITEDISABLEDIRSYNC, OmitFeatures, SQLITEOMITAUTOMATICINDEX, SQLITEOMITBETWEENOPTIMIZATIO0, SQLITEOMITBTREECOUNT, SQLITEOMITLIKEOPTIMIZATIO0, SQLITEOMITLOOKASIDE, SQLITEOMITOROPTIMIZATIO0, SQLITEOMITQUICKBALANCE, SQLITEOMITSHAREDCACHE, SQLITEOMITXFEROPT, Options,} \newline
 *SetAutoVacuum, AutoVacuumOff, AutoVacuumO0, SetCacheSize, StandardCacheSize, LowerCacheSize, HigherCacheSize, LockingMode, ExclusiveLock, NormalLockingMode, PageSize, StandardPageSize, LowerPageSize, HigherPageSize, HighestPageSize & Performance  \hh
 SS3 & Berkeley DB C~ & havecrypto, havehash, havereplicatio0, haveverif1, havesequence, havestatistics, diagnostic, pagesize, ps1k, ps4k, ps8k,ps16k, ps32k, cachesize, cs32mb, cs16mb,cs64mb, cs512mb & Performance \hh
 SS8 & Apache Server & Base, HostnameLookups, KeepAlive,EnableSendfile, FollowSymLinks, AccessLog,ExtendedStatus, InMemor1, Handle & Performance  \hh
 SS10 & Roll Sort &  spouts, maxspout, sorters, emitfreq, chunksize, messagesize & Throughput \hh
\end{tabular}
\label{t:datasetEnterprise}
\end{table*}

\section{Detailed Results}
\label{s:resultsexplained}

In this section, we show the detailed results for the six data sets listed in Table~\ref{t:dataset}. In all cases, we depict the results pictorially with the x-axis as the normalized value of each CV $p_i$ and the y-axis as the normalized value of the respective health index ($h_i$ for CV's). Each dot represents a health index $h_i$ computed at available data-point $p_i$. Depending on the availability and variance of data, some graphs have denser dots than others. The shape of the graphs shows the performance functions as discovered by the CHI model. We explain these observations in the following.

\subsection{Discovering the Influence of configuration objects}
\label{s:result-csg}

Fig.~\ref{f:sampleData-csg} shows a small subset of the CSG data-set $\DD_{test}$ with all configuration parameters normalized to fall in the range 0..1. Each row represents an input configuration file  ($c_n$'s) and the columns correspond to the configuration objects ($P_{m}$'s). CV names ($P$'s) are given in the header row and the last column refers to the observed output metric ($O_n$'s, in this case, performance expressed as bits/sec).  

The results in Fig.~\ref{f:resultsData-csg} show  the final `discovered' health index $h_{nm}$'s in each cell $\{n,m\}$ for various configuration object values $p_{nm}$ based on the above regression solution. CHI computes the $\gamma$'s \& $\eta$'s for each configuration object $P_m$'s to satisfy the objectives explained earlier and computes the overall health index of the configuration file (\textit{H}'s). The last two columns of Fig.~\ref{f:resultsData-csg} show that the computed \textit{H}'s  is closely related to the observed metric (last column $O$'s).  During this discovery phase, the minimum MSE achieved was around 0.0128 after 500 iterations.

\begin{figure}[htb]
\vspace{-6pt}
\centering
\includegraphics[width=0.9\columnwidth]{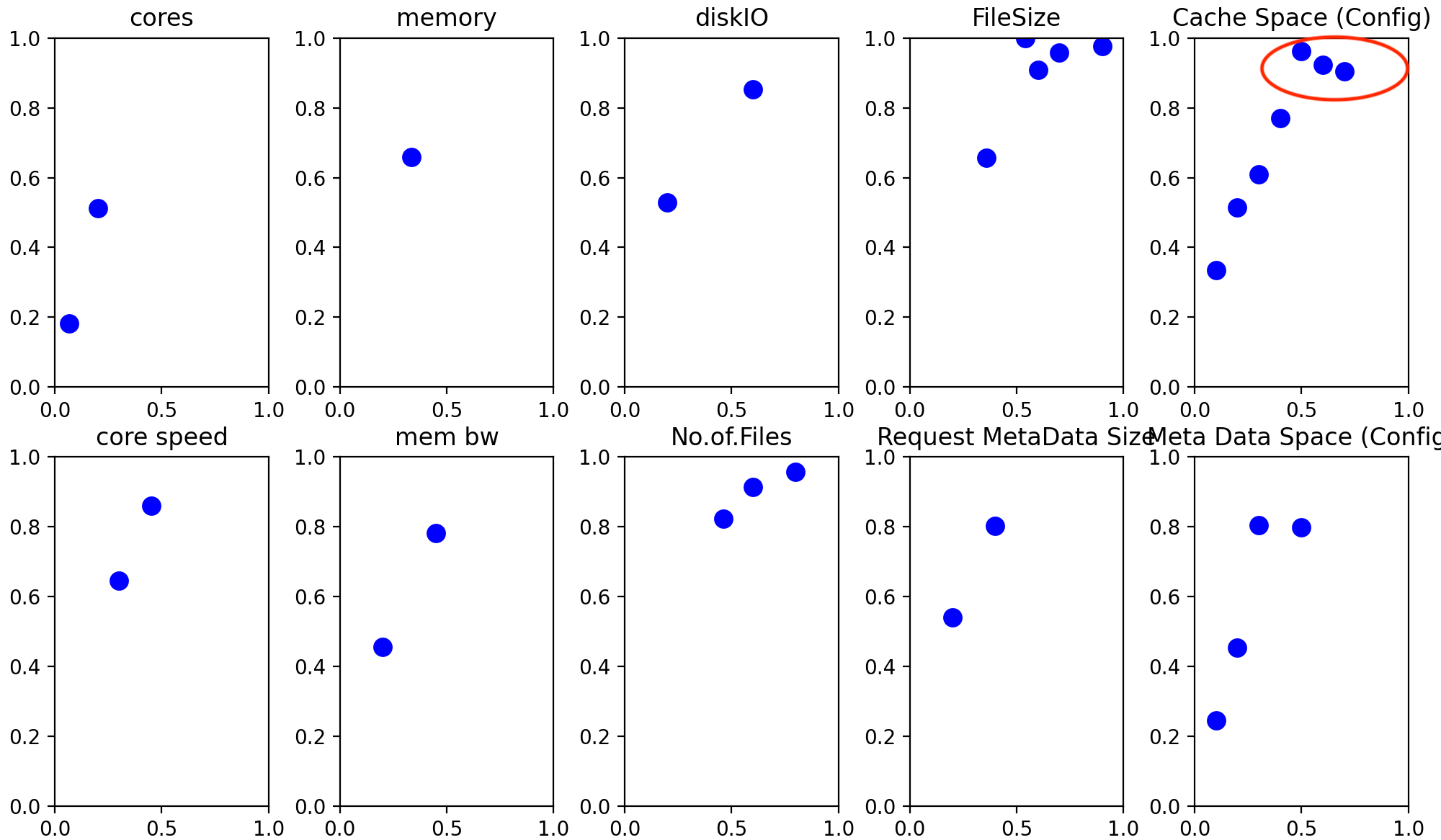}
\vspace{-6pt}
\caption{Results: HI metrics for CSG}
\label{f:results-hiValues-CSG}
\vspace{-12pt}
\end{figure}

After regressing through the data-set to achieve the desired minimum MSE, CHI correlates the individual configuration object values $p_{nm}$'s and their respective $h_{nm}$'s and determines the ``influential behavior'' of each of the CVs ($P_m$'s). With the discovered $\gamma$'s \& $\eta$'s, CHI can \textit{build a picture} of how each of these CVs affect the final outcome $O_n$. This relationship is shown in Fig.~\ref{f:results-hiValues-CSG}. In this figure, the x-axis shows the normalized values of each CV (shown as the label above sub-graph) and the y-axis is the normalized value of the respective health index ($h_i$ for $P_i$), and the name of the CV given above the sub-graphs.

These figures demonstrate that CHI can discover the behavior with respect to each CV including the strength of the influence, the cut-off point of diminishing return $P^{(mode)}_{m}$, and rate of decay afterward. The graphical results in Fig.~\ref{f:results-hiValues-CSG} can be visualized by the user to understand how different CVs influence the configuration \textit{H} (and in turn the service behavior).

We examine these graphical results closely and show that the results are indeed supported by our in-depth study of CSG domain~\cite{mass2019sondur}. For example, it is seen that the CSG performance is unimodal with respect to the Cache Space and Meta-Data Space, i.e., there is a threshold beyond which any further increase is detrimental to the system performance.  Our earlier CSG research work~\cite{mass2019sondur} supports this as it showed that allocating excessive cache space (i.e., blindly throwing resources at the problem) does not help. The CSG needs to perform background tasks such as garbage collection, data eviction to Cloud, data-refresh, etc. Allocating excessive data cache buffer (see  sub-graph in Fig.~\ref{f:results-hiValues-CSG}) can hurt these background processes, taking additional time to examine the data in the cache and reduce performance. Similar findings on  meta-data space configuration is supported by our CSG work in that excessive meta-data space allocation will trigger large metadata operations which in turn takes time, CPU, and memory resources and reduces performance.

\subsection{Behavior with "New" Configurations}
\label{s:result-newconfig}

We use the discovered values, i.e. outcome of the optimization objective ($\gamma_m$'s \& $\eta_m$'s)  to determine the $H^{new}$ of a set of a new (unseen) configuration file. We use the $2^{nd}$ part of the split empirical data set $\DD_{test}$ to validate the CHI. The  $H^{new}$ is computed using the validation model in the CHI framework (marked shaded in Fig.~\ref{f:framework}). Note that the computation of new $H^{new}$ does not dependent on the compute unit or input from experts or regression logic, because the characteristics of various $P_m$'s is already discovered and stored in the CHI repository. Fig.~\ref{f:resultsData-csg} shows the computed $h_{nm}$ and $H^{new}$ of the new configuration files. The last two columns in Fig.~\ref{f:resultsData-csg}  show that the newly computed $H^{new}$ is closely related to the observed metric $O^{new}$s (i.e. the true value). This set of results demonstrates that CHI can reasonably determine the probable behavior of the service (i.e observable metric $O$'s)  of the new configuration files using the $\gamma$'s \& $\eta$'s discovered earlier.  The MSE and variance for different train/test ratio data-set for various systems  is given in Table~\ref{t:dataset}.

\subsection{CHI for "BitBrains" Data}
\label{s:results-bb}

\begin{figure}[htb]
%\vspace{-6pt}
\centering
\begin{tabular}{c}
\subfigure[{Sample FastStorage (RND 500) Configuration File}]{
\includegraphics[width=0.99\linewidth]{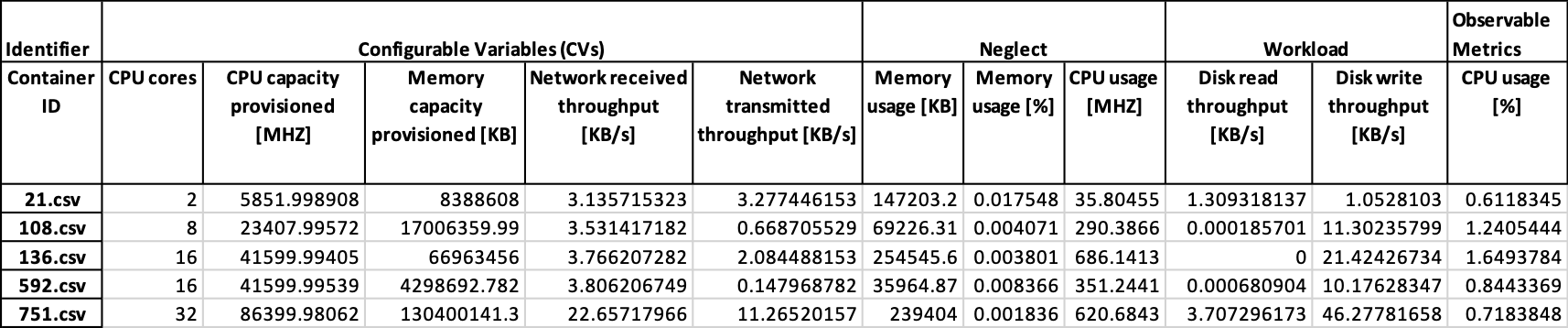}
\label{f:sampleData-bb}}  \\
%\vspace{6pt}
\subfigure[{Results: HI metrics for FastStorage (RND 500)}]{
\includegraphics[width=0.9\linewidth]{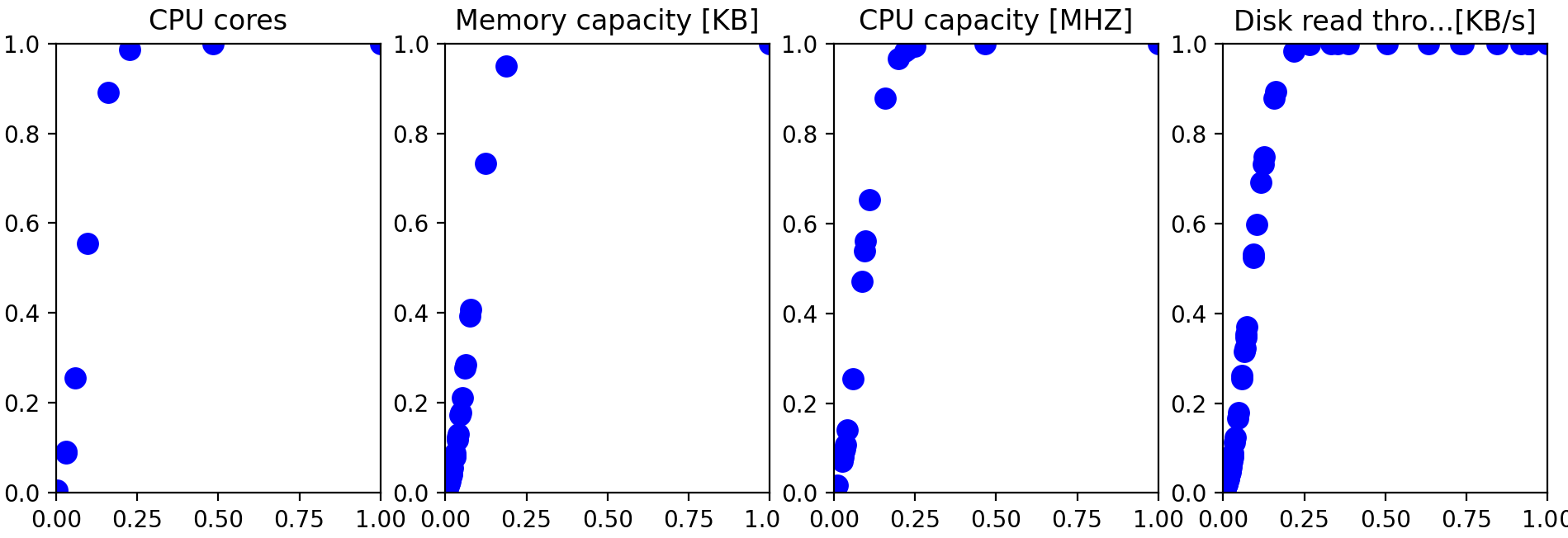}
\label{f:bbData-results}} 
\end{tabular}
%\vspace{-6pt}
\caption{Empirical Data and \textit{BitBrains} (RND 500) Results}
\label{f:figure-BB}
%\vspace{-24pt}
\end{figure}

In the absence of an explicit throughput measure, we quantify CPU utilization as an observable metric ($O$'s) and the remaining attributes as CVs ($P$'s). The latter can be changed and allocated differently for the various VMs. In the absence of any further information, we identify each VM as a unique configuration with its associated compute, memory, disk IO, and network resources. A sample of raw data-set used in our studies is given in Fig.~\ref{f:sampleData-bb}. Using this data, we restate the above research question as: \textit{Quantify the influence of various CVs of the VM on the CPU utilization in Bitbrain data-center}. 

\begin{table}[htb]
\vspace{-6pt}
\centering
\caption{Description of BitBrains GWA-T-12 Rnd traces.}
\begin{tabular}{|p{1.50in}|p{1.0in}||p{1.50in}|p{1.0in}|}
 \hline
\textbf{Data-set Variables} & \textbf{Usage} & \textbf{Data-set Variables} & \textbf{Usage} \hh
VM Container ID  & Identifier &
 &  \hh
CPU Capacity (MHz) & \multirow{5}{4em}{Configuration Variables} & Timestamp &  \multirow{5}{4em}{Ignored} \\
No. of CPU Cores  &  & Memory Usage (MB) & \\
Network Data Rcvd. (KB/s) & & Memory Usage (\%) & \\
Memory Capacity (MB) &  & CPU Usage (MHz) & \\
Network Data Transmit (KB/s) & & & \hh
Disk Read Throughput (KB/s) & \multirow{2}{4em}{Workload Characteristics} & CPU Usage (\%) & Observed Behavior \\
Disk Write Throughput (KB/s) & & \hh
\end{tabular}
\vspace{-6pt}
\label{t:bbdata-set}
%\vspace{-12pt}
\end{table}

Since the detailed time-series for each VM setting is not of interest here, we first compute the average value of every parameter for each VM. Given the long length of the trace, the averages should be quite reliable. The results indicate that a few VMs are outliers, with either almost no resource usage in spite of significant resource allocation, or very large  resource usage of one type (e.g., VMs that only do very intensive IO). We filtered out all zero value records as this would make the average resource usage so tiny that the entire exercise will be useless. After filtering, we normalized the data-set and used it for input to the CHI model.  The results are shown in Figs.~\ref{f:bbData-results}, with each sub-graph showing the influence of a CV on the observable metric. In all the graphs, the x-axis denotes the normalized values of each CV $P_i$ (shown by a label above the graph) and the y-axis is the normalized value of the chosen output metric ($O_i$), namely the CPU utilization.

A set of sample health index graph for few CVs of Apache server is shown in Fig.~\ref{f:figure-BB}. With limited insight into the this  data-set, we can theorize that performance as a function of the four CVs shown (namely number of CPU cores, memory size, CPU speed, and the disk IO rate) shows a familiar diminishing returns behavior with saturation. This is exactly what we would expect from a basic domain knowledge of computer architecture and IO modeling. For example, the overall CPI (cycles per instruction) for a workload depends on many factors and thus decreasing only one parameter (e.g., core CPI or access latency) will provide the kind of behavior we see in these graphs.
%The graphs (for a subset of CVs) confirm that the CPU utilization (and hence the overall throughput) reaches saturation quickly as a function of configured resources  such as the amount of memory provisioned, and disk/network throughput. 
Note that all VMs simply share the available SAN capacity (in terms of disk space and IO throughput), and network capacity. Also, since multiple VMs share the same underlying physical resources, a VM configuration can saturate quickly without yielding additional performance benefits, as the bottleneck can lie elsewhere.

\subsection{CHI for Enterprise data sets}
\label{s:results-nair}

The key results from the CHI model were summarized earlier in Table~\ref{t:dataset} (see rows for SS2,3,8,10). With the exception of Roll-Sort, which we discuss shortly, the MSE and its variance are quite low consistently, from about 1.7\% to 6.2\%.  Furthermore, the learning time ranged between 10 to 25 seconds for all these data sets. {\em These results substantially surpass the prediction results in the literature using these data sets both in terms of accuracy and time.} For example, ref.\cite{guo2013variability} uses incremental random samples with steps equal to the number of configuration options (features) of the system. They show rather unstable predictions with a mean prediction error of up to 22\%, and a standard deviation of up 46\%. Ref.~\cite{siegmund2015performance} discusses a technique that learn predictors for configurable systems with low mean errors, but the variance in the predictions could be very large; in particular, in half of the results for the Apache Web server predictions, standard deviation was up to 50\%. Also, the learning time is reported to be  1-5 hrs depending on the data-set.

\begin{figure}[htb]
\subfigure[{Results: HI metrics for SQL Lite Configuration}]{
\includegraphics[width=0.9\linewidth]{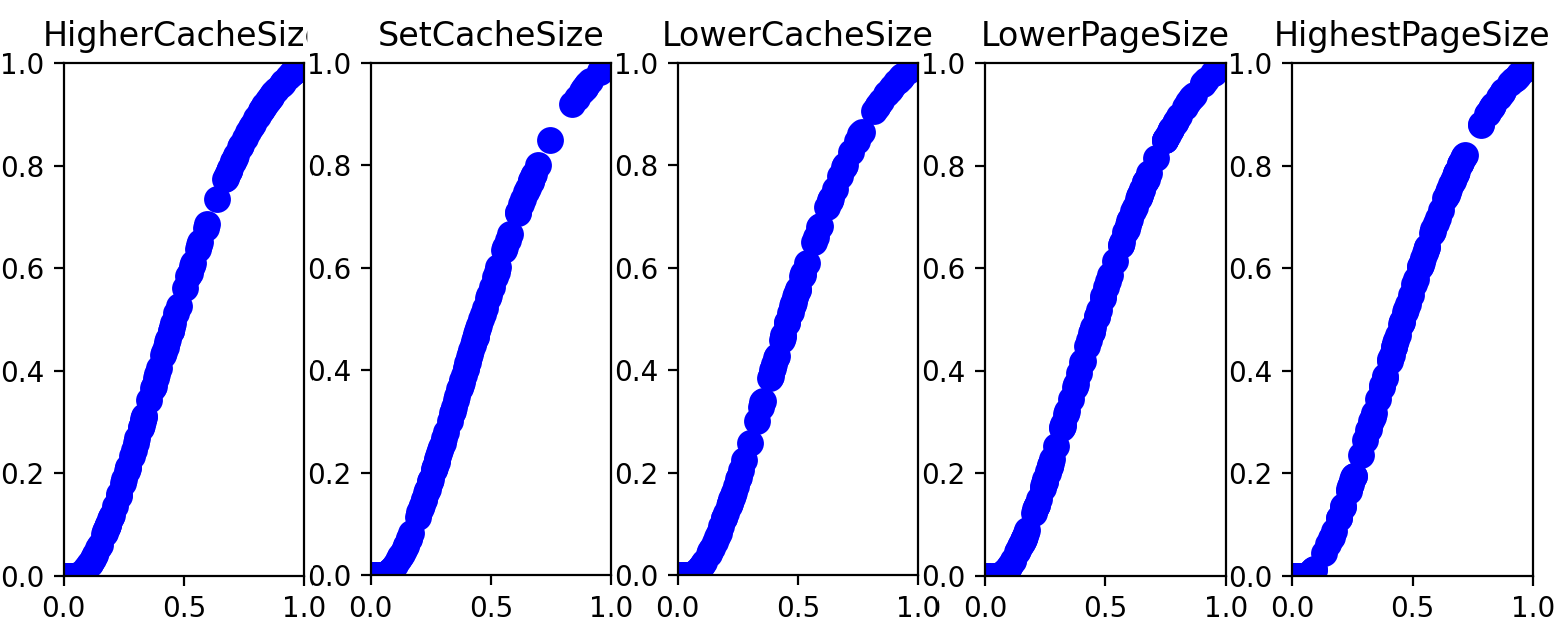}
\label{f:sqlliteResults}}
\end{figure}

\begin{figure}[htb]
\caption{Results: HI metrics for Roll-Sort Configuration }
\includegraphics[width=0.9\linewidth]{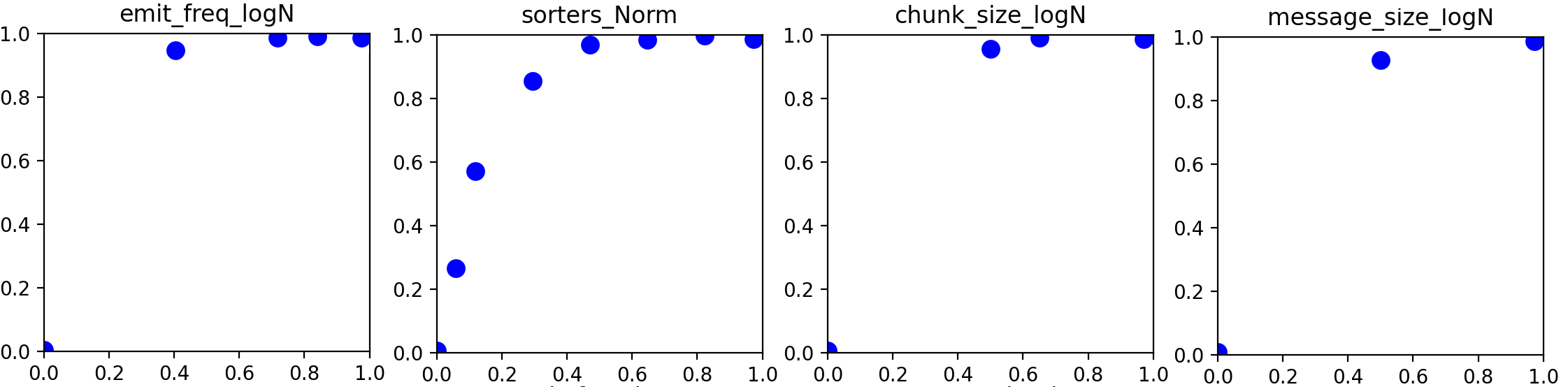}
\label{f:rs6dc3Results}
%\vspace{6pt}
\end{figure}

Before discussing the results in Fig.~\ref{f:sqlliteResults} pm SQL Lite, we note an important point about its configuration settings. Like most real-world databases, SQL Lite has a large number of configuration parameters, but many of them do not have much influence on the performance. The model used in Ref.~\cite{nair2017using} had several unexplained options compared to SQL Lite documentation~\footnote{SQL Lite doc. at \url{ https://www.sqlite.org/c3ref/c_config_covering_index_scan.html} \& \url{https://www.sqlite.org/c3ref/c_dbconfig_defensive.html}.}. While we cannot speak directly about this data-set, it appears (based on our deep understanding of how relational databases operate), that these additional parameters (which represent some minor options to be turned on/off) should not have a strong influence on the SQL lite performance. We thus decided to exclude them in our CHI modeling is shown in Fig.~\ref{f:sqlliteResults}. The excluded CVs are marked as an \textit{italicized} font in Table.~\ref{t:datasetEnterprise} and CVs considered by CHI is marked as a normal font (after marker *).

Finally, we show the CHI models for the Roll-Sort (SS10) workload in Fig.~\ref{f:rs6dc3Results}. Unlike other workloads, which represent complex applications, Roll-sort is merely a sorting algorithm and has only six CVs, but it is unclear what's special about and whether this is an external sort. The CHI model shows that the influence of several CVs saturates at certain values and any further increase in the resource (e.g. No.of sorters, chunk size) does not result in better performance. However, it appears that the data here is very noisy, perhaps influenced by the IO subsystem.

Ref.~\cite{nair2017using} mentions that for several software systems in their study, the configuration spaces are far more complicated and hard to model. They color code these hard-to-model system as yellow and red (Fig. 1 in Ref.~\cite{nair2017using}). Further, they state that applying the state-of-the-art technique by Guo at al.~\cite{guo2013variability} on these software systems showed the error rates of the generated predictor up to 80\%.
Using the data-set for the same systems used by Ref.~\cite{guo2013variability} (See Table II \& III), CHI showed a substantial improvement in error rate as shown in Table~\ref{t:datasetComparision1}. With 50\% $\DD_{train}$ \& 50\% $\DD_{test}$ for these data sets, CHI achieved an MSE for SQL Lite at 6.2\%, for Berkeley DB C: 4.17\%, and for Apache Server: 3.71\%. CHI can outperform in most cases since the objective is to discover the influence of individual CVs rather than focus on building a detailed performance model. Additionally, CHI does not depend on the sampling techniques which are again data dependent.

\begin{table}[htb]
%\vspace{-12pt}
\caption{Error Rates of Enterprise Data-set}
\centering
\begin{tabular}{|p{0.25in}|p{0.80in}|P{1.00in}|P{1.20in}|} \hline
{Code}  & {System} & {CHI Error Rate} & {Error Rate in Ref.~\cite{guo2013variability}}\hh
 SS2 & SQL Lite &  6.20\% $\pm$ 3.11\% & 7.2\% $\pm$ 4.2\% \hh
 SS3 & Berkeley DB C~ & 4.17\% $\pm$ 2.19\% & 6.4\% $\pm$ 5.7\%  \hh
 SS8 & Apache server & 3.71\% $\pm$ 2.1\% & 9.7\% $\pm$ 10.8\%  \hh
\end{tabular}
\label{t:datasetComparision1}
\end{table}

\section{Discussion: Segmented Regression (MARS \& LARS)}
\label{s:discussionsSegmentedRegression}

The influence of individual CVs on the health index can be complicated~\cite{zhu2017bestconfig} and is generally not linear. Yet much of the data-driven behavior characterization attempts to fit linear or piece-wise linear segments to the observations. In particular, if we have $m$ predictor variables ($X=\{x_i\}$, i $\in 1 \cdots M$) (CVs in our study) and observed output ($Y$) (performance in our study), a typical assumption is a linear relationship along with a normally distributed error term $\epsilon$ with zero mean and variance $\sigma^2$:
\EQ
Y = X \beta + \epsilon  \text{, where }
\epsilon = \mathcal{N}(0,\sigma^2)
\label{e:genericLinear}
\EN

{\bf Linear Regressions -- OLS, Ridge, Lasso:} Such regression algorithms aim to estimate $\hat{\beta}$ (the unknowns) such that some measure of overall error is minimized. For example,  the  ordinary least square (OLS) regression minimizes the sum of squares of residuals to achieve the unbiased estimate:
\EQ
L_{OLS}(\hat{\beta}) = 
\sum_{i=1}^{N} (y_i - x_i{\beta})^2 
\text{, and  minimize }
 \left( \frac{|Y-X\beta|^2}{n} \right)
\label{e:OLS}
\EN

Other algorithms such as Ridge or Lasso regression try to reduce variance at the cost of introducing some bias. For example, Lasso regression adds the constraint $\sum_{j=1}^{M} (|\beta_j| < t)$ where $t$ is a given threshold. Lasso has a parsimony property~\cite{efron2004least,tibshirani2013lasso}: for any given constraint value $t$ , only a subset of the predictor variables (i.e. $x_i$'s) have nonzero values. i.e. many predictor variables \textit{can} have a zero thereby suppressing their contribution to the output $Y$. That is, in the configuration problem at hand, these algorithms try to ``suppress'' the contribution of some CVs. 

Further, Ref.~\cite{Wang2019} argues that machine learning based analytical models, though have been shown to work very well in some specific scenarios, do not consider the domain specific practical factors such as non-linear multi-threading overhead or JVM GC activities, which are very related to soft resource allocation and can significantly degrade server efficiency. Our evaluation supports this statement with empirical results, as given in section~\ref{s:resultsSegmentedRegression}.

{\bf Multivariate Adaptive Regression Spline (MARS):} MARS ~\cite{friedman1991multivariate,milborrow2014earth} is a technique for deriving simple multi-segment models from the data. It can be viewed  as an extension of a linear model that automatically models non-linearities and interactions between variables by combining hinge functions of the form $\pm max(0, x- K)$, (where $K$ is a constant). MARS builds a linear model of the form:
\EQa
\hat{y} = \hat{f(x)} = \sum_{i=1}^{k} c_iB_i(x) 
\label{e:mars}
\ENa
where the predicted value ($\hat{y}$) is a sum of coefficient ($c_i$) and basis function ($B_i(x)$). Our investigation revealed that a greedy model like MARS uses brute force to derive the above parts  of the model ($c_i$'s \& $B_i$'s), and the hinge function cut-off points ($K$).  Though the MARS model can yield good results for predicting new outcomes,  an uninformed model like MARS for CHI has little regard for the physics of the problem and may behave in unexpected ways such as eliminating certain important CVs or put in hinge points (i.e., change in slope) at unexpected places or increase/decrease slope in unexpected ways. For example, instead of showing a steady diminishing-returns property that applies in almost any situation with increasing resources, MARS may as well use a line segment with a larger slope on the higher end!

{\bf Least Angle Regression (LARS):} LARS~\cite{tateishi2010nonlinear,Plan2016,efron2004least} produces a full piece-wise linear solution path to  a non-linear relationship between predictor variables $x_i$'s and output $y$. LARS algorithm is similar to forward step-wise regression, but instead of including variables at each step, the estimated parameters are increased in a direction equiangular to each one's correlations with the residual. In section~\ref{s:resultsSegmentedRegression}, we show the limitations of LARS in discovering the influence of the CVs on the performance, wherein the algorithm ignores important CVs though there is a wider variance of such data.

In designing a solution, the ``goodness'' is often defined in terms of prediction accuracy, but parsimony is another important criterion since simpler models provide better insight into the $X \Rightarrow Y$ relationship~\cite{efron2004least}. However, we believe that this tradeoff (i.e., more segments implying better accuracy) is introduced somewhat artificially by restricting the model to linear segments which ignore the physics of the problem. Instead, our approach is to find a nonlinear function that shows the desired characteristics (e.g. smoothness, diminishing returns, complexity related loss in performance, etc.) without splitting into more \& more segments. We show that such an approach not only correctly captures the expected behavior of the system, it is also less complex.

\subsection{Results: Segmented Regression (MARS \& LARS)}
\label{s:resultsSegmentedRegression}

\begin{table*}[htb]
\begin{minipage}{\textwidth}
%\vspace{-12pt}
\caption{Results from Segmented Regression (MARS \& LARS) for 80\% $D_{train}$ \& 20\% $D_{test}$}
\centering
\begin{tabular}{|p{0.5in}|p{1.5in}|p{1.5in}|P{0.65in}|P{0.65in}|P{0.65in}|} \hline
{Code}  & {CVs considered by MARS} & {CVs ignored by MARS} & {MARS MSE} & {LARS MSE} \hh
{CSG}  & {FileSize, No.of.Files, Cores} & {Core Speed, Memory, Mem BW, Disk IO, Cache Space, Meta-Data Space, Req.MetaData Size} & {0.0015} & {0.0054} \hh
{BB}  & {NW transmitted, NW received, Memory, Disk Read} & {CPU cores, CPU capacity, Disk Write,}  & {0.0253} & {0.0327} \hh
{SQL Lite}  & {SetAutoVacuum, AutoVacuumO, AutoVacuumO0, SetCacheSize, PageSize, HigherCacheSize, ExclusiveLock,  StandardPageSize} & {StandardCacheSize, LowerCacheSize, LockingMode, NormalLockingMode,  LowerPageSize, HigherPageSize, HighestPageSize} & {0.0262} & {0.0260} \hh
{Berkeley DB C}  & {have crypto, diagnostic, ps1k, ps4k, ps8k, ps16k, ps32k, cs16mb, cs512mb} & {have hash, have replicatio0, have verif1, have sequence, have statistics, pagesize, cachesize, cs64mb} & {0.0166} & {0.0170} \hh
{Apache}  & {EnableSendle, KeepAlive, Handle, InMemor1} & {Base, HostnameLookups, AccessLog, ExtendedStatus, FollowSymLinks} & {0.0138} & {0.0147} \hh
%{Roll Sort}  & {maxspout, messagesize,  sorters} & {emitfreq, chunksize} & {0.0236} & {0.0236} \hh
\end{tabular}
\label{t:otherResults}
%\vspace{-18pt}
\end{minipage}
\end{table*}

Although the segmented regression algorithms (MARS and LARS) can do a good job of fitting the data and thereby yield superior prediction accuracy within the range covered by the data, a blind faith in data is particularly troublesome for physical systems where we do understand many things about reasonable vs. anomalous behavior. For example, the  artificial data fitting by these algorithms often runs counter to sensible behavior, such as showing a higher slope with more resources (i.e., superlinear behavior) where generally one would expect diminishing returns and hence a flattening trend. Even worse, these algorithms may kick out the important CVs and keep the irrelevant ones since they do not have any insight into the nature of individual predictors.

\begin{figure*}[!ht]
\vspace{-6pt}
\centering
\begin{tabular}{cc}
\subfigure[{CSG}]{
\includegraphics[width=0.45\linewidth]{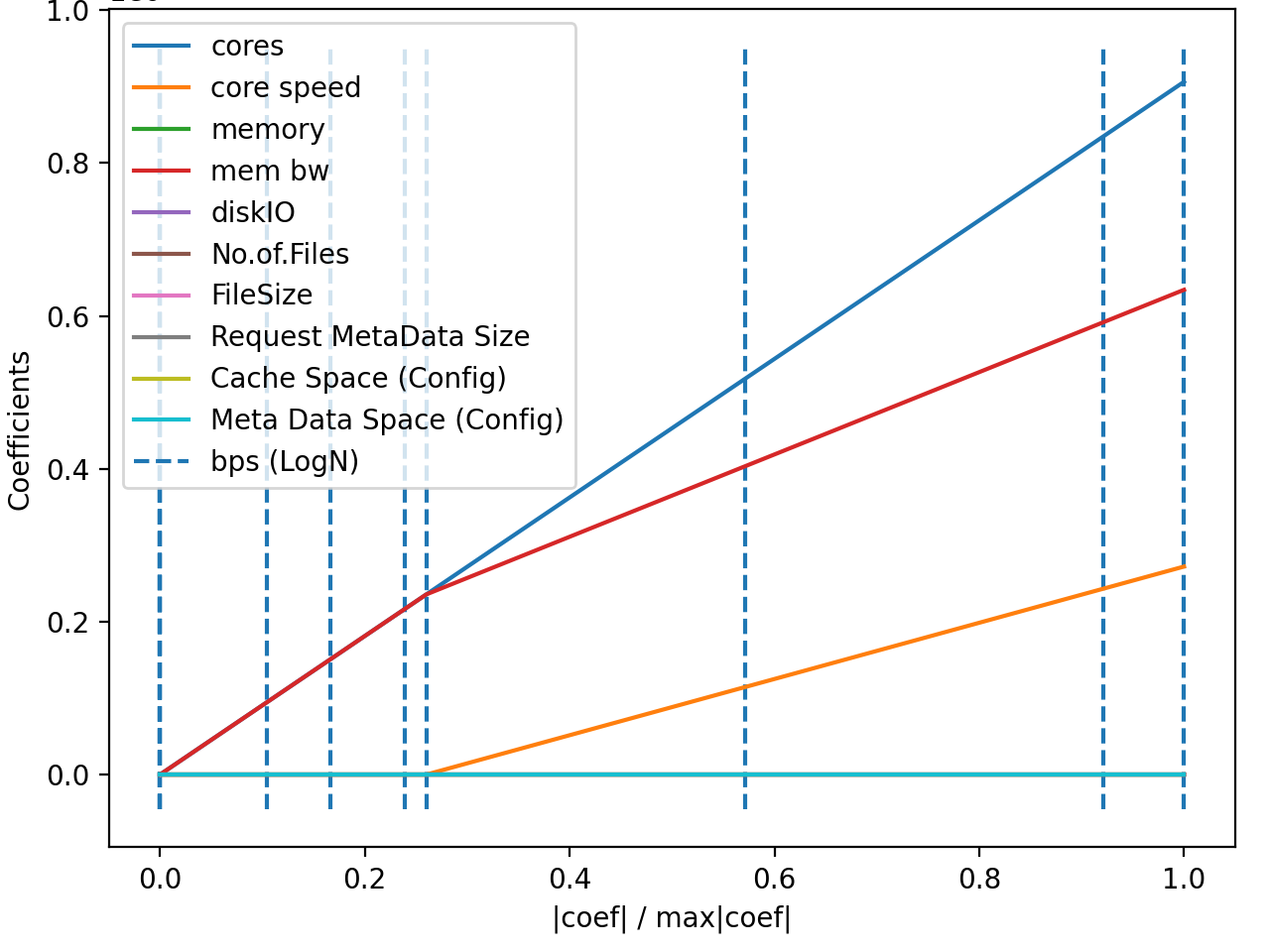}
\label{f:lars-csg}}  &
%\vspace{6pt}
\subfigure[{BB}]{
\includegraphics[width=0.45\linewidth]{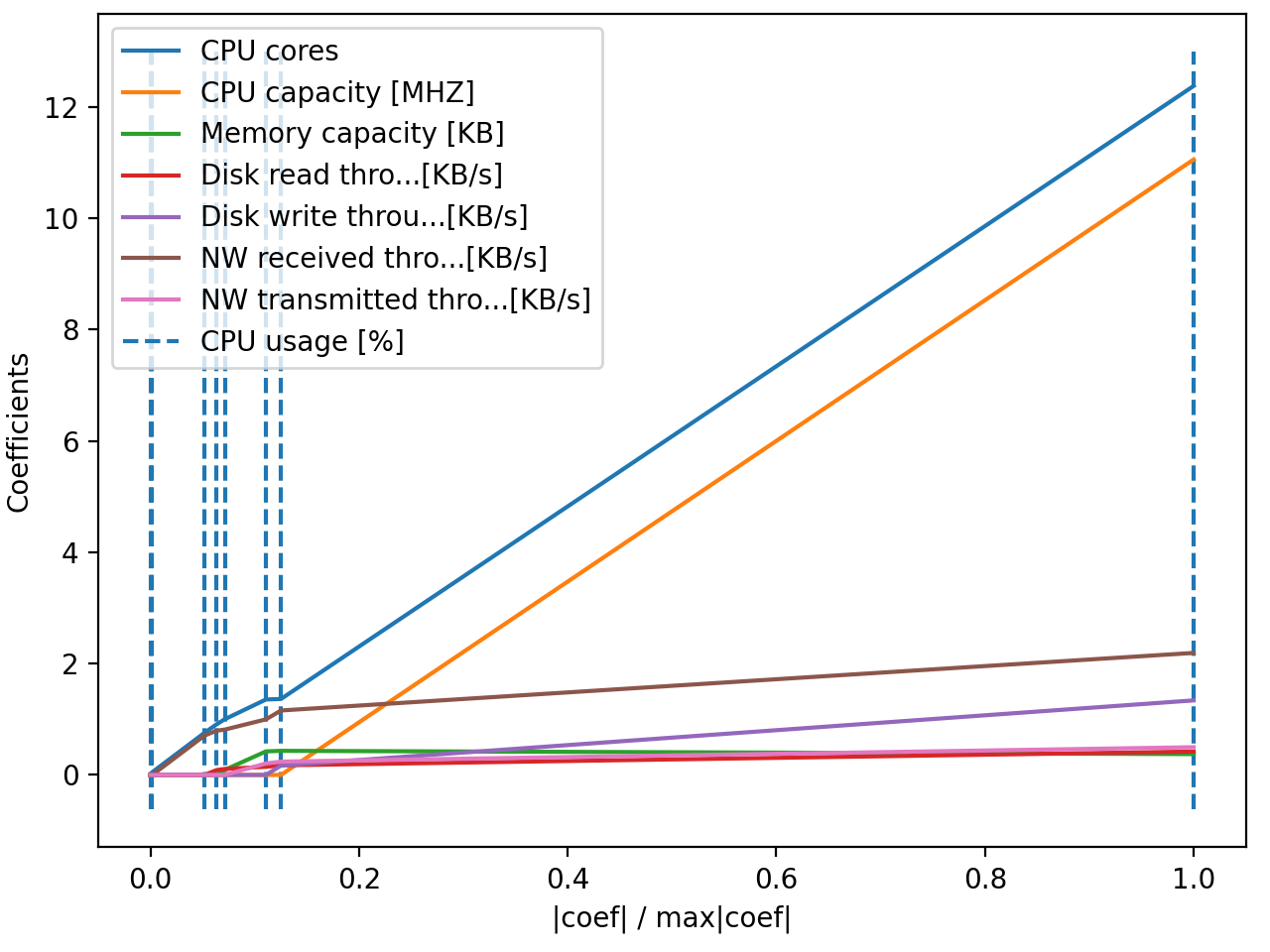}
\label{f:lars-bb}} \\
\subfigure[{Apache}]{
\includegraphics[width=0.45\linewidth]{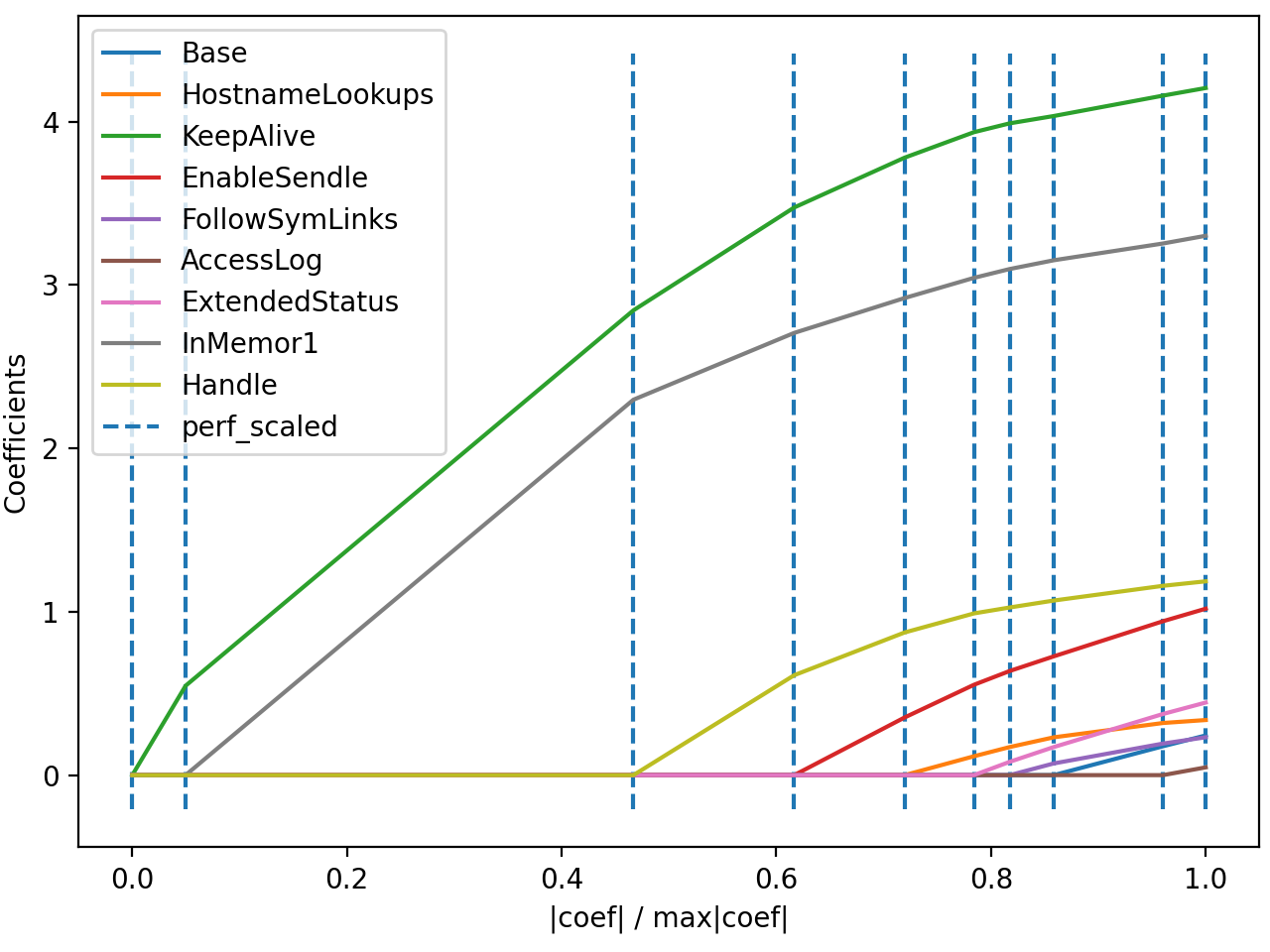}
\label{f:lars-apache}}  &
%\vspace{6pt}
\subfigure[{BDBC}]{
\includegraphics[width=0.45\linewidth]{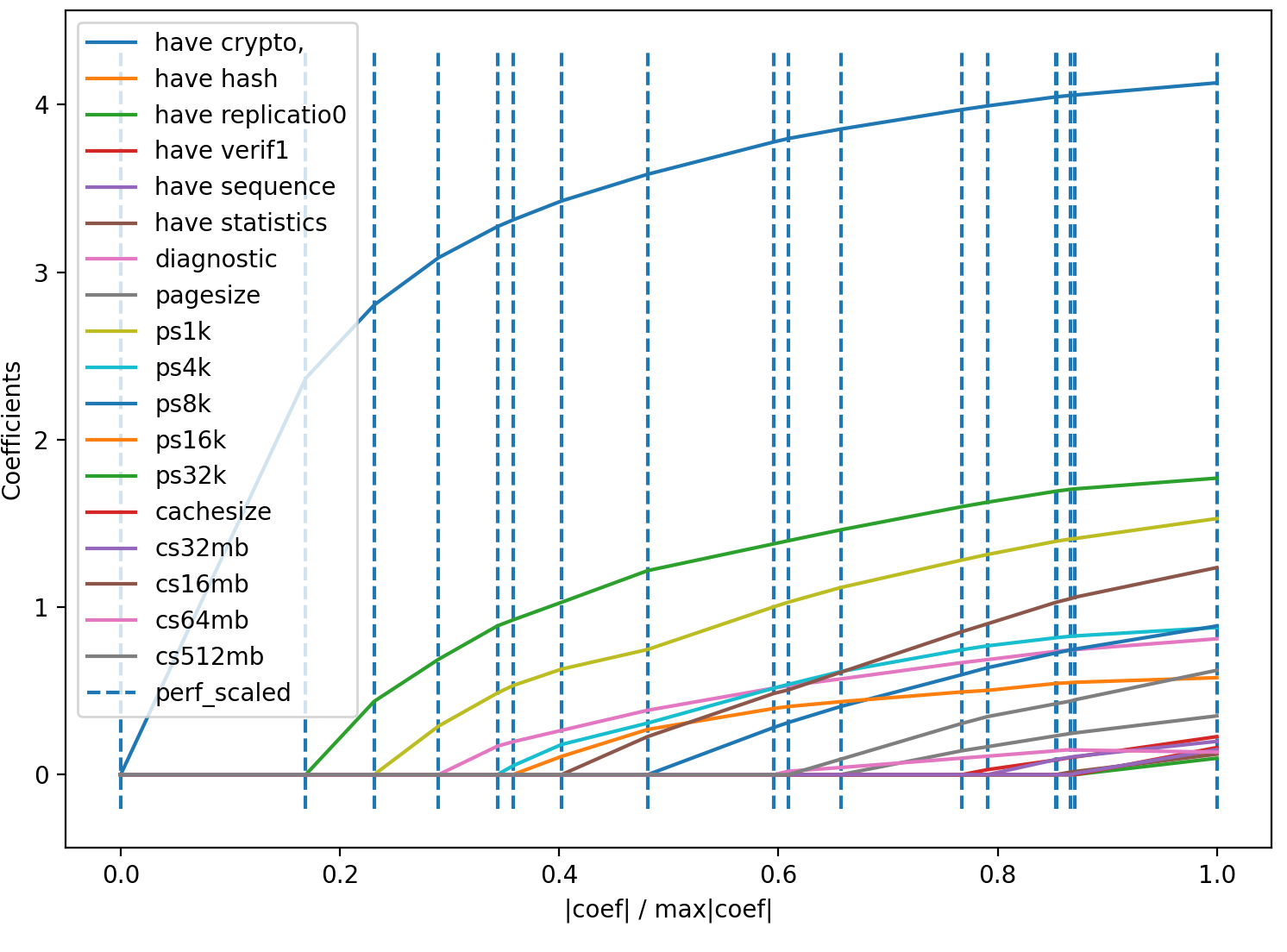}
\label{f:lars-bdbc}} \\
\subfigure[{SQL Lite}]{
\includegraphics[width=0.45\linewidth]{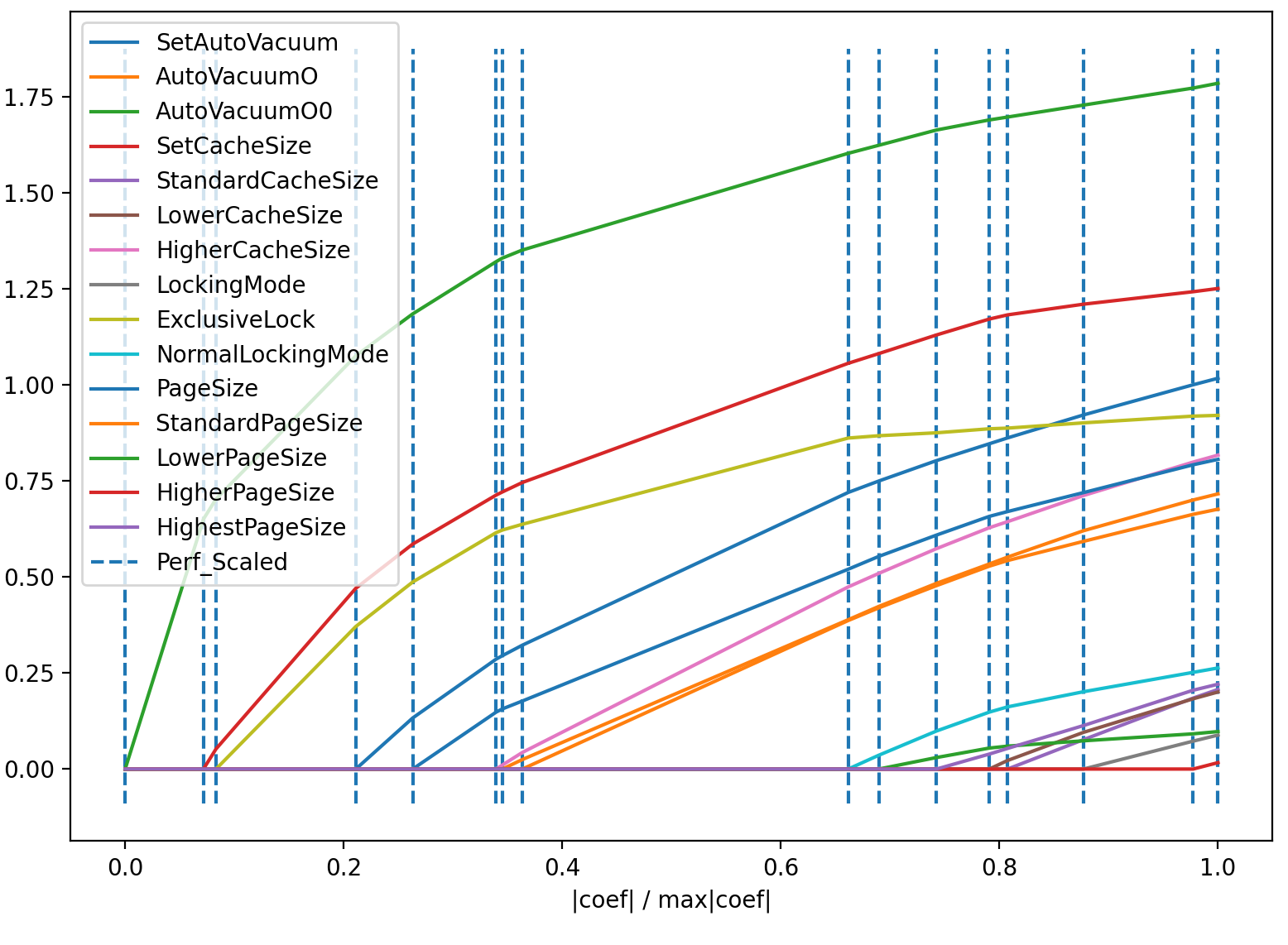}
\label{f:lars-sqllite}}  &
\subfigure[{Roll Sort}]{
\includegraphics[width=0.45\linewidth]{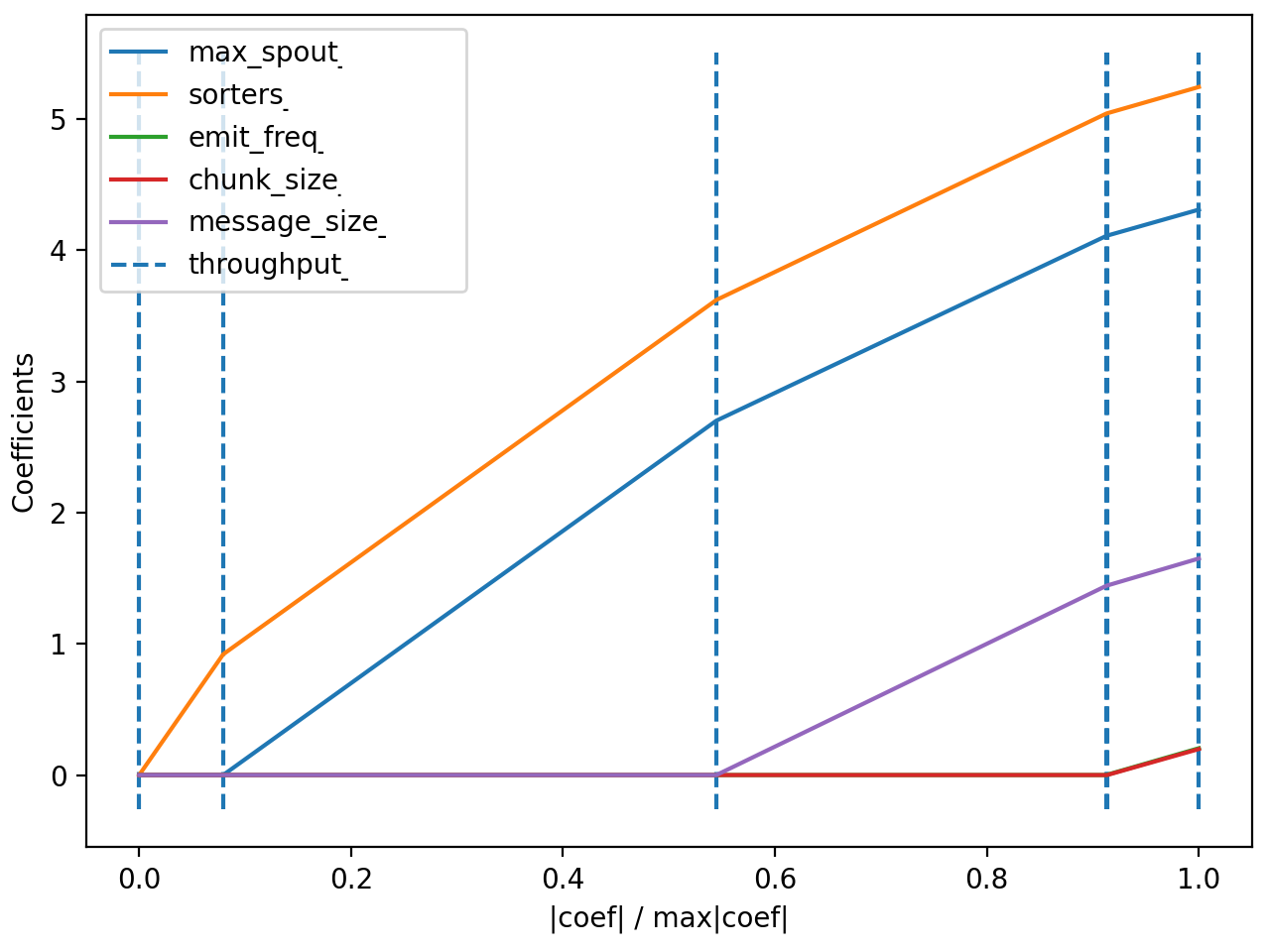}
\label{f:lars-rollsort}}
%\vspace{6pt}
\end{tabular}
\vspace{-6pt}
\caption{Results: LARS Coefficients (weights) for above data sets}
\label{f:figure-LARS}
%\vspace{-24pt}
\end{figure*}

%MARS
For example, MARS uses a brute force algorithm to regress over a CV $P_i$ to reach the best possible MSE before considering the next CV $P_j$. This is evident from the results as shown in Table.~\ref{t:otherResults}, wherein MARS ignores several CVs for all the domains. In our work involving CSG, we (as  experts who have significant experience with it) can confidently say that the ignored CVs (CacheSpace, Meta-DataSpace, Req.MetaData Size, Memory etc.) have a dominant bearing on the performance of the system. As has been noted in our earlier work~\cite{Cloud2019Sondur,mass2019sondur}, although File size and No. of Files are prominent workload characteristics that do have a bearing on the performance, but they are not the primary components that can be isolated from the rest. Similar observations for BitBrains VM components show that MARS ignores most of the CVs and the performance prediction is solely based on two components (Network \& Disk). VM performance experts tend to argue that compute capacity (CPU cores, CPU core speed) influences the performance heavily.

%LARS
We show the results from LARS in Fig.~\ref{f:figure-LARS} for different domains, where   the x-axis shows  the normalized values of the  CV settings ($p_i$'s) and the y-axis shows the normalized values of the performance of the system. Fig.~\ref{f:lars-csg} is the LARS output for CSG illustrating that performance is heavily dependent on only three components (cores, core speed, memory bandwidth). As systems people, we know that the performance is not dominated by one or two components, but is dependent on a balance between compute, memory, disk IO \& workload. Similar results are evident in Fig.~\ref{f:lars-bb}, where the VM performance is linearly dependent on two prime components (CPU cores \& CPU capacity[MHz]), largely ignoring the rest of the CVs. This is again in contrast with VM domain knowledge -- as basic architecture knowledge would indicate, the compute resource does not have a linear relationship with performance. Instead, the performance depends on the overall CPI (cycles per instruction) which is impacted by cache and memory path latencies. Finally, although MARS \& LARS use a similar approach for segmented regression (i,e. converting a non-linear relationship into a series of linear regions), we see that they yield drastically different results, which too is troubling and indicates a dissociation from the physics of the problem.

% Variance of data-set...
% cores	core speed	memory	mem bw	diskIO	testtype1	No.of.Files.Orig	No.of.Files	File Size Orig	FileSize	Request MetaData Size	Cache Space (Config)	Meta Data Space (Config)	Log-Class	Put-Class	Put-Speed	Put-Speed-Log	Cache Space (LogN)	MetaData Space (LogN)	Log Space (LogN)	Bps	BPS-Class	bps (LogN)
%0.4%	0.5%	0.0%	1.5%	3.9%	4.9%	9.0%	1.6%	2.9%	2.6%	1.0%	3.2%	1.0%	1.5%	1.8%	0.6%	0.2%	4.3%	1.3%	2.1%	4.4%	3.9%	1.2%
%
%CPU cores	CPU capacity [MHZ]	Memory capacity [KB]	Disk read thro...[KB/s]	Disk write throu...[KB/s]	NW received thro...[KB/s]	NW transmitted thro...[KB/s]	CPU usage [%]
%1.3%	1.2%	0.3%	1.6%	1.0%	1.0%	0.3%	4.0%

This  study validates our argument that a blind belief in data-driven models or direct application of a well-versed algorithm is not only counterproductive but risks interpreting the results with no attention to the dynamics of the system under study.

To encourage future extensions to CHI and future study in configuration subject areas, we  share the full data-set used in Table~\ref{t:dataset}, the python implementation of CHI, MARS \& LARS, and the full set of results  at {\url{https://www.kkant.net/config_traces/CHIproject}}.

\section{Current State of the Art and Challenges}
\label{s:stateofart}
 
Many domain specific articles speak about challenges pertaining to the configuration (or resource allocation) of networks~\cite{fernandes2019comprehensive,kakarla2020finding,fogel2015general}, compute units or storage~\cite{anderson2002hippodrome}, operating systems~\cite{config-errors}, applications~\cite{siegmund2015performance,Wang2019}, Cloud ~\cite{Masanet2020,moradi2021online}, etc.
A prominent approach in the literature on configuration settings has been the {\em performance influence model (PIM)} that captures the relationship between CVs and the performance ~\cite{siegmund2015performance,Calotoiu2016,nair2017using}. PIM is almost entirely dependent on model training using available performance data  and does not reflect or exploit any domain knowledge concerning either the relationships or the limitations that go beyond the range of training data.  PIM like approaches look at the statistical influence of the configuration values and do not consider the design and architecture but are learned from observations~\cite{zhang2015performance}. A pure statistical model simply fits the data to a model but does not provide any insights into whether or why the real behavior is compatible with the statistical observations. In contrast, CHI aims at identifying  the dominant properties of the CVs and quantify their parameters using the data. 

Xu et al.~\cite{xu2015hey} report that the Apache server has more than 550 parameters and many of these parameters have dependencies and correlations, which further worsens the situation. Reference ~\cite{siegmund2015performance,nair2017using} narrows this down to  only nine  CVs configuration options  at \url{http://tiny.cc/3wpwly}, but no rationale is given. In particular, the thread-pool size of Apache Server is not considered but is reported to be critical in Wang et al.~\cite{Wang2019}. {\em We believe that such issues can substantially benefit by exploiting domain knowledge of the administrators instead of simply depending on the data, which could be misleading or inadequate.}

Probability based approaches for finding optimum configurations such as ConEx~\cite{krishna2020conex} are a variation of the PIM model that probabilistically sample the configuration space and then generate a machine learning (ML) model to predict an outcome (usually performance). However, the contribution of individual configuration parameters on the outcome is not modeled. Variability aware models proposed by Guo et al.~\cite{guo2013variability} work on boolean CVs (being set true/false), but it is well known that arbitrary Boolean functions of this form simply cannot be learned~\cite{zhang2015performance}. As discussed by Zhang et al. ~\cite{zhang2015performance}, we show that performance functions are not arbitrary, but rather structured, hence can be potentially learned effectively.

Velez et al.\cite{velez2020configcrusher} \& Ha et al.~\cite{ha2019deepperf} observed that the influence of configuration parameters on performance is highly variable, i.e., some options are highly influential while others have little or no impact on the performance. Such performance variations have made it very challenging to predict the performance of an application running in the Cloud environment. CHI postulates the performance as a function of each important configuration parameter based on the domain knowledge (e.g., monotonic with diminishing returns) and thus reduces the configuration space and the data requirements to quantify the behavior.

Xu et al.\cite{xu2016early} in their study of various application configurations reveal that about 4.7\%–38.6\% of the critically important CVs do not have any early checks and  thereby cause severe impact on the system’s behavior. Xu focuses the study on CVs related to the system’s Reliability, Availability, and Serviceability (RAS). This concept is in line with our approach in that CHI rules out unimportant CVs (explained later as $L_{un}$) and service behavior is expressed with a measurable health index metric).
Wang et al.~\cite{Wang2019} show that liberal allocation of a CV (i.e  DB pool size) can lead to performance degradation. Further, their study shows the importance of considering the practical factors such as non-linear effect of resource allocation. CHI supports this observation and discovers the non-linearities in CVs.

%Cloud Computing Journal:
In Cloud environments, Zaman et al.~\cite{Zaman2013} show that VM provisioning depends heavily on resource allocation which in turn affects economics and bidding process (e.g. $VM_1$ with 1x2-GHz CPU, 8GB memory, 1TB disk vs. $VM_2$ with 2x2-GHz CPU, 16GB memory, 2TB disk;).  Using practical data from a Cloud Storage environment~\cite{Cloud2019Sondur}, we show that CHI can help understand the effect of resource allocation. Zhu et al.~\cite{zhu2017bestconfig} demonstrate the difficulty and infeasibility of the configuration tuning problem  using common machine learning model-based methods. Wei et al. ~\cite{wei2015towards} and Moradi et al.~\cite{moradi2021online} highlight the complexity of allocating multiple resource types in a study of heterogeneous resource allocation in Cloud VMs. By understanding how multiple resource types (e.g., number of CPU cores, disk size, etc.) affect the performance/workload, CHI can aid users in reducing the monetary costs by choosing the right heterogeneous and economical resource allocation, thus be also cost-efficient.

%Sampling techniques

%other tools
It is also worth noting that although there are many Configuration Management Tools (e.g., CFEngine, Puppet, Ansible,  etc.~\cite{onnberg2012software}), their job is only the application of provided settings to multiple resources consistently and ensuring that certain given relationships hold.

\section{Conclusions}
\label{s:conc}

The behavior of all cyber systems depends on a set of configuration variables (CVs) which if set improperly could result in a variety of problems including sub-optimal performance. %In CHeSS~\cite{icin2020sondur}, we designed a framework to capture the "health index" (H) of a service with respect to attributes (or dimensions) such as performance, security, availability, etc. CHeSS depended on human inputs for assignment of weights to quantify the influence of CVs which is unscalable in a large system.  
In this paper, we present a performance related Configuration Health Index (CHI) framework that can  {\em quantify} the  contribution of individual CVs towards the overall performance of the service. We evaluate CHI using a model-driven approach that exploits both the domain knowledge and the available data. We demonstrate the applicability of CHI using data sets from the state-of-art systems and our study of Cloud Storage Gateway. In all cases, we demonstrate that CHI can learn the influence of CVs on service performance and accurately predict the behavior for new (yet unseen) configuration settings. We show that our approach works better than a pure data-driven characterization and can give a better {\em a priori insight} into the influence of the CVs on the system performance. We believe that CHI provides a substantial improvement over the state of the art and can be broadly applicable to wide range of configuration management problems. We also demonstrate the dangers of the pure data driven models in that they might predict a wrong trend or eliminate important configuration variables. An approach that uses data judiciously along with the domain knowledge based constraints can address this problem.

\section*{Acknowledgements}
\label{s:ack}
We appreciate the support and active participation of Girisha Shankar (Ph.D student) from Indian Institute of Science, Bengaluru, India and Dr. Slobodan Vucetic of Temple University. The discussions with them were highly valuable in devising the solution and added to the techniques presented in the paper.

\bibliographystyle{acm}
\bibliography{referencesA, myReferences,  config_mgmt, config_cvss}

\end{document}